\documentclass[aps,prb,twocolumn,showpacs,amsmath,amssymb,superscriptaddress]{revtex4-2}
\usepackage{graphicx}
\usepackage{CJK}
\usepackage{color}
\usepackage[colorlinks,bookmarks=false,citecolor=blue,linkcolor=red,urlcolor=blue]{hyperref}
\usepackage{multirow}
\usepackage{ulem}
\usepackage{tabularx}
\usepackage[nounderscore]{syntax}
\graphicspath{{figs/}}
\usepackage{subfigure}
\usepackage{float}

\newcommand{\be}{\begin{equation}}
\newcommand{\ee}{\end{equation}}

\newcommand{\sxz}[1]{\textcolor{blue}{\textbf{[sxz: #1]}}}

\begin{document}
\begin{CJK*}{UTF8}{gbsn}
\title{Numerical instability of non-Hermitian Hamiltonian evolution }

\author{Xu Feng}
\affiliation{Beijing National Laboratory for Condensed Matter Physics, Institute of Physics, Chinese Academy of Sciences, Beijing 100190, China}
\affiliation{School of Physical Sciences, University of Chinese Academy of Sciences, Beijing 100049, China}

\author{Shuo Liu}
\affiliation{Institute for Advanced Study, Tsinghua University, Beijing 100084, China}

\author{Shi-Xin Zhang}
\email{shixinzhang@iphy.ac.cn}
\affiliation{Beijing National Laboratory for Condensed Matter Physics,
Institute of Physics, Chinese Academy of Sciences, Beijing 100190, China}

\author{Shu Chen}
\email{schen@iphy.ac.cn}
\affiliation{Beijing National Laboratory for Condensed Matter Physics,
Institute of Physics, Chinese Academy of Sciences, Beijing 100190, China}
\affiliation{School of Physical Sciences, University of Chinese Academy of Sciences, Beijing 100049, China}

\begin{abstract}
The extreme sensitivity of non-Hermitian Hamiltonians exhibiting the non-Hermitian skin effect (NHSE) has been extensively studied in recent years with well-established theoretical explanations. However, this sensitivity is often overlooked in numerical simulations, leading to unreliable results. In this work, we reexamine Hatano-Nelson and symplectic Hatano-Nelson models studied in previous work [Kawabata \textit{et al}., Phys. Rev. X 13, 021007 (2023)], and compare their results with our high-precision calculations. We systematically investigate inaccuracies in physical results arising from numerical instability during diagonalization and non-Hermitian Hamiltonian evolution. We find that these instabilities arise from a large condition number that scales exponentially with system size due to the NHSE, signaling strong non-normality.
Strikingly, a reliable spectrum alone is shown to be insufficient for accurate non-Hermitian evolution, while the reliability of wave functions plays a more critical role. Our findings underscore the necessity of evaluating the condition number to ensure the validity of numerical studies on systems with NHSE, implying that some prior numerical findings in this area may require careful reexamination.

\end{abstract}
\maketitle
\section{Introduction}
In quantum mechanics, Hermiticity is required for closed systems to ensure real observables and probability conservation. However, when a system exchanges energy, particles, or information with its environment, it undergoes non-unitary evolution described by the master equation. In certain cases, this non-unitary dynamics can be effectively described by a non-Hermitian Hamiltonian (NHH), which has garnered significant interest in recent years.
Numerous novel phenomena associated with NHHs have been discovered, including the non-Hermitian skin effect (NHSE) and exceptional points (EPs), which have no counterparts in Hermitian systems \cite{ashida2020non,GBZ,BiorthogonalBBC,AnatomySkinModes,Slager,SvdNHSE,CNHSE,EPs,HuEP}. Notably, NHSE and EPs have been experimentally observed in various quantum platforms \cite{xiao2020,Coldatom1,Coldatom2,DigitQuantumComputer}. Furthermore, non-Hermitian descriptions have also been applied to a wide range of non-conservative classical systems, leveraging the formal equivalence between the single-particle Schr\"odinger equation and the classical wave equation. Owing to their high controllability, many non-Hermitian phenomena have been verified in electrical, photonic, acoustic, and mechanical systems \cite{acoustic,mechanical,Helbig2020,ReciprocalNHSECircuit,SuLuHong,Guo2024}. 

Mathematically, EPs
and NHSE arise from the unique properties of non-Hermitian matrices. EPs are special points in the parameter space where two or more eigenvalues and their corresponding eigenvectors coalesce, rendering the matrix defective \cite{EPs}. Systems near EPs exhibit extreme sensitivity to parameter variations \cite{ashida2020non}.
Beyond EPs, NHSE represents another key mechanism of extreme sensitivity. Two hallmark signatures of NHSE are the accumulation of a macroscopic number of eigenstates near the system's boundary and a dramatic change in the energy spectrum under different boundary conditions \cite{non-bloch,WindingSkin,WindingSkin2,boundarysensitive}. This phenomenon is closely linked to the spectral instability of non-Hermitian matrices. In general, if a non-Hermitian matrix $H$ is perturbed by an arbitrary matrix $\Delta$, the resulting spectral shift is not necessarily bounded by the norm $\lVert \Delta\rVert$~\cite{ashida2020non,pseudospectra}.
This instability poses significant challenges for numerical computations. Previous studies have shown that the spectrum obtained via numerical diagonalization can be unreliable due to extreme sensitivity to lattice size and numerical precision \cite{auxiliaryGBZ,AlgorithmSpectrum}. To address this issue, the Generalized Brillouin Zone (GBZ) theory has been developed for analytically characterizing the spectrum of one-dimensional non-Hermitian systems in the thermodynamic limit \cite{GBZ,non-bloch,auxiliaryGBZ}.

Nevertheless, numerical calculations of non-Hermitian matrices remain unavoidable. For instance, the GBZ theory is inapplicable to disordered, high-dimensional, and interacting non-Hermitian systems. Moreover, recent research has shifted beyond spectral analysis to focus on the dynamics of non-Hermitian systems \cite{NHEvoQuasiDisorder,NHEvoAA,liDisorderInducedEntanglementPhase2023,Turkeshi,LinHuNHEvo,MultipolesNHSE,EEManyBodyHN,longhiNHEvo,NHEvoYangLee}. Our study focuses on potential numerical errors in non-Hermitian evolution. As anticipated, these errors can be highly sensitive to lattice size and numerical precision. We analyze this phenomenon through the lens of pseudospectra and demonstrate, using the Hatano-Nelson and symplectic Hatano-Nelson models, as studied in previous work, that numerical errors can lead to qualitatively incorrect physical results \cite{NHSEEPTKawabata,Schiro}.
Finally, we highlight the distinction between our work and previous studies, particularly Ref. \cite{auxiliaryGBZ}. While Ref. \cite{auxiliaryGBZ} emphasizes the sensitivity of the spectrum to lattice size and numerical precision, our work focuses on wave functions and non-Hermitian evolution. Notably, we show that even if the spectrum remains accurate within an acceptable error margin, the results of non-Hermitian evolution can still be significantly incorrect.

The rest of the paper is organized as follows. In Sec. \ref{SecKnowledgeNonNormal}, we introduce the basic knowledge of the non-normal matrices, which theoretically demonstrates that strong non-normality can induce spectrum instability and unconventional dynamics. Next, we analytically solve the Hatano-Nelson and symplectic Hatano-Nelson models in Sec. \ref{SecExamples}, whose condition number and pseudospectra are also obtained. Sec. \ref{SecNumericalErrorDiag} presents the energy spectra and wave functions of the Hatano-Nelson and symplectic Hatano-Nelson model with different numerical precisions, highlighting the numerical instability during diagonalization. 
Sec. \ref{SecNHEvo} provides the results of non-Hermitian evolution for both models at various numerical precisions, further revealing the instability of non-Hermitian evolution. In Sec. \ref{SecMisunderstanding}, we address a common misunderstanding, namely that a reliable spectrum alone does not guarantee the reliability of non-Hermitian evolution, while the accuracy of the wave functions is more critical. Finally, Sec. \ref{SecConclusion} concludes the paper and offers several perspectives for future work. Appendix \ref{AppendixNumericalAsymmetry} 
provides evidence that the observed numerical asymmetry arises from the $QR$ algorithm.
Appendix \ref{AppendixGaussianStateSimulation} introduces the Gaussian state simulation method and various observables of interest. In Appendix \ref{AppendixUEvo}, we monitor the evolution of the $U$ matrix to observe the manifestation of numerical errors directly. Appendix \ref{AppendixExtraNumericalRes} provides additional numerical data to further support the findings presented in the main text. The generalizations to the disordered and interacting systems are briefly illustrated in Appendix \ref{AppendixDisorderInteraction}.

\section{Fundamentals of the non-normal matrix}\label{SecKnowledgeNonNormal}
In this work, we consider the case of a non-Hermitian Hamiltonian, where $H\neq H^{\dagger}$. Specifically, we primarily focus on non-normal matrices, which demand $[H, H^{\dagger}]\neq 0$.
For any normal matrix, the spectral theorem guarantees the existence of a unitary transformation $U$ that diagonalizes the matrix, i.e., $H=U\Lambda U^{-1}$, where $\Lambda$ is a diagonal matrix representing the energy spectrum, and the columns of $U$ correspond to the eigenstates. However, if $H$ is non-normal but still diagonalizable, there exists an invertible (albeit non-unitary) matrix 
$V$ such that $H=V\Lambda V^{-1}$. Due to the non-unitarity of $V$, both left and right eigenvectors are required to construct a biorthogonal basis.
In extreme cases, such as at EPs, the non-Hermitian Hamiltonian becomes non-diagonalizable, meaning its geometric multiplicity is lower than its algebraic multiplicity. In such cases, the Hamiltonian can only be transformed into the Jordan normal form.

For non-normal matrices, the spectrum can be highly unstable. Before proceeding, we emphasize that all norms used throughout this paper refer to the $2$-norm. For a vector $x$, the 2-norm is  $\lVert x\rVert=(\sum_j|x_j|^2)^{1/2}$, and for a matrix $A$,  $\lVert A\rVert=\text{max}_{x}\dfrac{\lVert Ax\rVert}{\lVert x\rVert}=s_{\text{max}}(A)$, where $s_{\text{max}}(A)$ is the largest singular value of $A$. The singular values of $A$ are the square roots of the eigenvalues of $A^{\dagger}A$. 
The discussion now turns to spectral instability, starting with the eigenvalue equation $H|\psi^{R}_j\rangle=E_j|\psi^{R}_j\rangle$ and its left-eigenvector counterpart $\langle\psi^{L}_j|H=\langle\psi^{L}_j| E_j$. If we perturb $H$ by adding an extra term $H(g)=H+g\Delta$ with $\lVert\Delta\rVert=1$, we can analyze the spectral shift using perturbation theory for small $g$.  
The perturbed eigenvalue equation is given by $H(g)|\psi^{R}_j(g)\rangle=E_j(g)|\psi^{R}_j(g)\rangle$, with the expansions $E_j(g)=E_j+gE_{j}^{(1)}+g^2E_{j}^{(2)}+\cdots$ and $|\psi^{R}_j(g)\rangle=|\psi^{R}_j\rangle+g|{\psi^{R}_{j}}^{(1)}\rangle+g^2|{\psi^{R}_{j}}^{(2)}\rangle+\cdots$. The first-order correction to the eigenenergy is $E^{(1)}_j=\langle\psi^L_j|\Delta|\psi^R_j\rangle/\langle\psi^L_j|\psi^R_j\rangle$. For non-normal matrices, even if $\lVert \langle\psi_{j}^{L}|\rVert$ and $\lVert|\psi^R_j\rangle\rVert$ are finite, the denominator $\langle\psi_j^L|\psi_j^R\rangle$ can be arbitrarily small. For instance, $|\psi^R_j\rangle$ and $|\psi^L_j\rangle$ of the Hatano-Nelson model may localize at opposite ends of the system, leading to a large $E_{j}^{(1)}$ and, thus, spectral instability. In contrast, for normal matrices, where $\langle\psi_j^L|=\langle\psi_j^R|$, the perturbative correction remains bounded.

A useful measure of non-normality and spectral instability is the condition number, defined as $\text{cond}(V)=\lVert V\rVert\cdot\lVert V^{-1}\rVert$. It can be proved that $\text{cond}(V)=s_{\text{max}}(V)/s_\text{min}(V)$, where $s_{\text{max}}(V)$ and $s_{\text{min}}(V)$ are the largest and smallest singular values of the matrix $V$, respectively. Immediately, for a normal matrix, $V$ is unitary, so $\text{cond}(V)=1$. As we will show, the condition number can be extremely large for a highly non-normal matrix.
Another insightful way to characterize spectral instability is through pseudospectra \cite{pseudospectra}, defined in several equivalent ways. Here, we list two commonly used definitions:  

(i) The $\varepsilon$-pseudospectrum $\sigma_{\varepsilon}(H)$ of a matrix $H$ is the set of complex numbers $z\in\mathbb{C}$ satisfying $\lVert(z-H)^{-1}\rVert>\varepsilon^{-1}$. 

(ii) For any $\varepsilon$, the $\varepsilon$-pseudospectrum $\sigma_{\varepsilon}(H)$ consists of all eigenvalues of perturbed matrices $H+\Delta$, where $\Delta$ is a complex random matrix with 2-norm less than $\varepsilon$. 

In the first definition, the matrix $(z-H)^{-1}$ is known as the resolvent of $H$ at $z$, which corresponds to the Green's function, where $z$ represents a complex frequency. It follows that $\lVert(z-H)^{-1}\rVert=[s_{\text{min}}(z-H)]^{-1}$, where $s_{\text{min}}(z-H)$ is the smallest singular value of $z-H$. This relation allows for the practical computation of pseudospectra via singular value analysis. The second definition provides a more intuitive interpretation. Pseudospectra satisfy the properties $\sigma_{\varepsilon_1}(H)\subseteq\sigma_{\varepsilon_2}(H)$, $0<\varepsilon_1\leq\varepsilon_2$ and $\bigcap_{\varepsilon>0}\sigma_{\epsilon}(H)=\sigma(H) $, meaning that the intersection of all pseudospectra recovers the exact spectrum. 

For a normal matrix, the $\varepsilon$-pseudospectrum is simply the union of the open $\varepsilon$-balls centered at the eigenvalues. More precisely, $\lVert(z-H)^{-1}\rVert=1/\text{dist}(z,\sigma(H))$, where $\text{dist}(z,\sigma(H))$ denotes the usual distance of a point to the spectrum $\sigma(H)$ in the complex plane. In contrast, for a diagonalizable but non-normal matrix $H=V\Lambda V^{-1}$, we have $(z-H)^{-1}=(z-V\Lambda V^{-1})^{-1}=V(z-\Lambda)^{-1}V^{-1}$. This yields the bound $\lVert(z-H)^{-1}\rVert\leq\lVert V\rVert\lVert V^{-1}\rVert\lVert(z-\Lambda)^{-1}\rVert=\text{cond}(V)/\text{dist}(z,\sigma(H))$. Thus, the $\varepsilon$-pseudospectrum satisfies $\{z|\ \text{dist}(z,\sigma(H))<\epsilon\cdot\text{cond}(V)\}$ (Bauer-Fike theorem \cite{pseudospectra}). For non-normal matrices, the condition number $\text{cond}(V)$ can be extremely large, leading to significant deviations between pseudospectra and spectra even for arbitrarily small perturbations. Obviously, this sensitivity affects not only the spectrum but also the eigenvectors. The condition number and pseudospectra have served as powerful tools for investigating NHSE \cite{NonnormalHamOkuma,pseudospectraNHSE1,pseudospectraNHSE2,scalingpseudospectraexponentiallysensitive,GeneralCriterionPseudoSpectraNHSE,quadraticBosonLindbladian1,quadraticBosonLindbladian2,quadraticBosonLindbladian3,Znidaric1,Znidaric2,boombustcycles}.

Besides the spectrum instability, the strong non-normality can induce counterintuitive dynamics \cite{pseudospectra}. Naively, considering the evolution of the $\lVert e^{-\mathrm{i}Ht}\rVert$, because $\lVert e^{-\mathrm{i}Ht}\rVert=\lVert Ve^{-\mathrm{i}\Lambda t}V^{-1}\rVert\leq \text{cond}(V)\cdot\lVert e^{-\mathrm{i}\Lambda t}\rVert$, if $H$ is a normal matrix, i.e. $\text{cond}(V)=1$, the evolution of $\lVert e^{-\mathrm{i}Ht}\rVert$ is determined by the spectrum of $H$ as expected. However, for a highly non-normal matrix of very large $\text{cond}(V)$, the transient dynamics can greatly deviate from the behavior of $\lVert e^{-\mathrm{i}\Lambda t}\rVert$. As we will demonstrate in Sec. \ref{SecNHEvo}, for the Hatano-Nelson model, $\lVert e^{-\mathrm{i}Ht}\rVert$ initially grows exponentially with time before stabilizing at late times, as predicted by the $\lVert e^{-\mathrm{i}\Lambda t}\rVert$. Interestingly, we will show that this sharp early-time growth of $\lVert e^{-\mathrm{i}Ht}\rVert$ is closely tied to the numerical instability of the non-Hermitian evolution. 

In summary, as shown in Table \ref{tab:NonNormalSummary}, both spectral perturbations and transient dynamics can be significantly amplified in non-Hermitian systems exhibiting the NHSE, indicating the potential for numerical instability. 

\begin{table}[t]
\caption{\label{tab:NonNormalSummary}
Summary of four key concepts for non-normal matrices: non-normality, condition number, spectral instability, and dynamical instability.}
\begin{ruledtabular}
\begin{tabular}{p{3.3cm} p{7.0cm} p{4.5cm}}
\textbf{Concept} &  \textbf{Key Expression} \\
\hline
Non-normality &
$[H, H^\dagger] \neq 0$  \\
Condition number &
$\text{cond}(V) = \lVert V \rVert \cdot \lVert V^{-1} \rVert = \dfrac{s_{\max}(V)}{s_{\min}(V)}$ \\
Spectral instability &
$ \{ z \,|\, \mathrm{dist}(z, \sigma(H)) < \varepsilon \cdot \text{cond}(V) \}$ \\
Dynamical instability &
$\lVert e^{-\mathrm{i}Ht} \rVert \leq \text{cond}(V) \cdot \lVert e^{-\mathrm{i}\Lambda t} \rVert$ \\
\end{tabular}
\end{ruledtabular}
\end{table}

\begin{figure}[htb]
\centering
\subfigure{
\includegraphics[height=3.6cm,width=8.4cm]{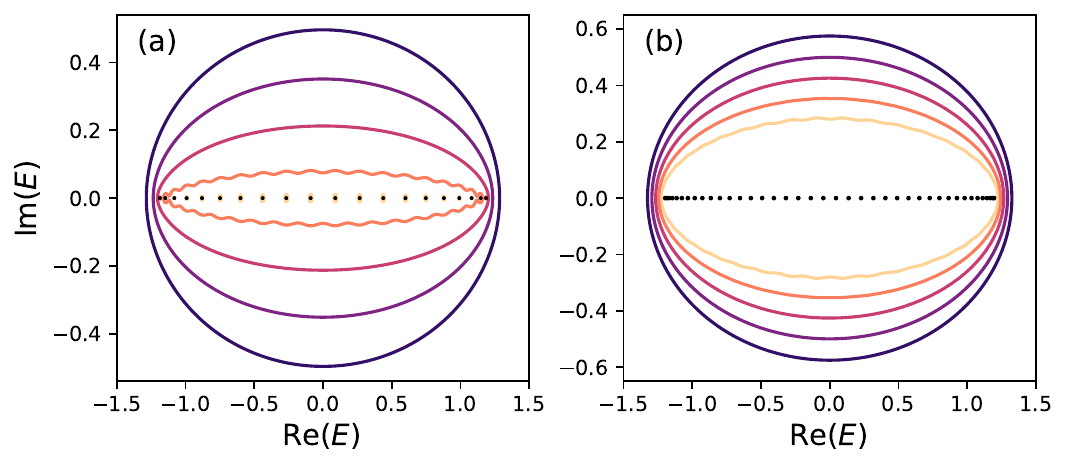}}
\caption{Boundaries of pseudospectra $\sigma_{\varepsilon}(H_{\text{HN}})$. Open boundary condition. $\gamma=0.8$ (a) $L=20$. From the lightest to the darkest color, $\varepsilon=10^{-10}\sim 10^{-6}$. (b) $L=40$.  From the lightest to the darkest color, $\epsilon=10^{-15}\sim10^{-11}$. In (a) and (b), the black solid points denote the exact energy eigenvalues under OBC. The pseudospectra is the interior enclosed by the boundary.}
\label{pseudospectraHN}
\end{figure}

\section{Illustrative examples}
\label{SecExamples}
\subsection{Hatano-Nelson model}
\label{HNmodel}
Here, we use the Hatano-Nelson model as an example to illustrate the spectrum instability in non-normal matrices.
The model is \cite{HNmodel}: 
\begin{equation}
H_{\text{HN}}=\sum_{j}(J+\gamma)c^{\dagger}_{j+1}c_{j}+(J-\gamma)c^{\dagger}_{j}c_{j+1}, 
\end{equation}
in which $c_{j}^{\dagger}$ and $c_j$ are the creation and annihilation operators of spinless fermions at site $j$. 
$J\pm\gamma$ is the hopping strength between neighboring sites. The non-reciprocity ($\gamma\neq0$) is the origin of NHSE and extreme sensitivity. Throughout the work,  we assume $J>\gamma$ ($J, \gamma\in \Bbb{R}$), and set $J=1$ as the energy unit.

The Hatano-Nelson model can be analytically solved under both periodic boundary conditions (PBC) and open boundary conditions (OBC).
Under OBC, the energy eigenvalues are given by 
$E_{\text{OBC}}^{(m)}=2\sqrt{J^2-\gamma^2}\text{cos}\theta_m$, $\theta_m=m\pi/(L+1)$, $m=1,2,...L$, whose spectrum is entirely real.  
In contrast, the energy eigenvalues under PBC are $E^{(m)}_{\text{PBC}}=2J\text{cos}k_m-\mathrm{i}2\gamma\text{sin}k_m$, $k_m=2m\pi/L$, $m=1,2,...L$, forming a loop encircling the OBC spectrum in the complex energy plane, a signature of the NHSE \cite{WindingSkin, WindingSkin2}.

The Hatano-Nelson Hamiltonian $H_{\text{HN}}$ under OBC can be transformed into a Hermitian matrix $A$ via the similarity transformation
\begin{equation}\label{Eq2}
H_{\text{HN}}=QAQ^{-1}=QU\Lambda U^{-1}Q^{-1}=V\Lambda V^{-1},
\end{equation}
in which $Q=\text{diag}\{r, r^2, ..., r^{L-1}, r^{L}\}$ with  $r=\sqrt{(J+\gamma)/(J-\gamma)}$. The corresponding Hermitian Hamiltonian is $A=\sqrt{J^2-\gamma^2}(c_{j+1}^{\dagger}c_j+\text{h.c.})$. Here, $\Lambda$ is a diagonal matrix representing the energy spectrum, and $V=QU$. The condition number of $V$ is given by $\text{cond}(V)=\text{cond}(Q)=r^{L-1}$, since multiplication by a unitary matrix leaves the condition number unchanged. It is important to note that $\text{cond}(V)$ is not unique due to the freedom in choosing eigenstates under a gauge transformation $G$. Thus, the relevant quantity to consider is $\text{cond}(VG)$. Fortunately, it has been proved that both the lower and upper bounds of $VG$ scale exponentially with system size $L$ as $r^{L-1}$ \cite{pseudospectraNHSE2}. In contrast, it can be readily proved $[H_{\text{HN}}, H_{\text{HN}}^{\dagger}]=0$ under PBC, namely $H_{\text{HN}}$ under PBC is normal, so the condition number under PBC is just $1$.  
In sum, for any nonzero $\gamma$, the condition number of the Hatano-Nelson model under OBC grows exponentially with the system size. In addition, with a fixed system size, the non-normality under OBC is greatly enhanced by increasing the non-reciprocity $\gamma$. On the contrary, the condition number of the Hatano-Nelson model under PBC is always $1$, independent of system size $L$ and non-reciprocity $\gamma$.

We now present the pseudospectra of the Hatano-Nelson model under OBC for different values of $\varepsilon$ in Fig. \ref{pseudospectraHN}. As shown in Fig. \ref{pseudospectraHN}(a), for $L=20$, $\gamma=0.8$, the $\varepsilon=10^{-10}$ pseudospectrum nearly coincides with the exact spectrum. However, for $\varepsilon$ ranging from $10^{-9}$ to $10^{-6}$, the boundary of the pseudospectrum progressively deviates from the exact spectrum. This suggests that under perturbations $\Delta$ with $\lVert\Delta\rVert<\varepsilon$, the perturbed spectrum $\sigma(H_{\text{HN}}+\Delta)$ differs significantly from the unperturbed spectrum $\sigma(H_{\text{HN}})$. As discussed earlier, the condition number of the Hatano-Nelson model under OBC grows exponentially with system size $L$, amplifying the sensitivity of the eigenvalues. In Fig. \ref{pseudospectraHN}(b), the system size is  $L=40$, and even for $\epsilon=10^{-15}$, the boundary of the pseudospectrum already exhibits a significant deviation from the exact spectrum. Notably, the default precision of floating-point numbers in most programming languages is approximately $10^{-16}\sim10^{-15}$, implying that numerical diagonalization within double precision is unreliable. Thus, the condition number provides a useful estimate of numerical reliability. For instance, in the Hatano-Nelson model with $L=20, J=1.0, \gamma=0.8$, the condition number is approximately $10^9$, suggesting that double precision is adequate. However, for $L=40$, the condition number increases to about $10^{18}$, rendering double precision potentially insufficient. In general, obtaining accurate numerical results for highly non-normal matrices requires improving numerical precision beyond a threshold $\varepsilon_{\text{th}}$, where the $\varepsilon_{\text{th}}$-pseudospectrum closely matches the exact spectrum within acceptable errors. When analytical results are unavailable, numerical convergence can be assessed by systematically increasing precision and observing the stabilization of the pseudospectrum, which utilizes the properties $\bigcap_{\varepsilon>0}\sigma_{\varepsilon}(H)=\sigma(H)$. 

\begin{figure}[htb]
\centering
\subfigure{
\includegraphics[height=3.6cm,width=8.4cm]{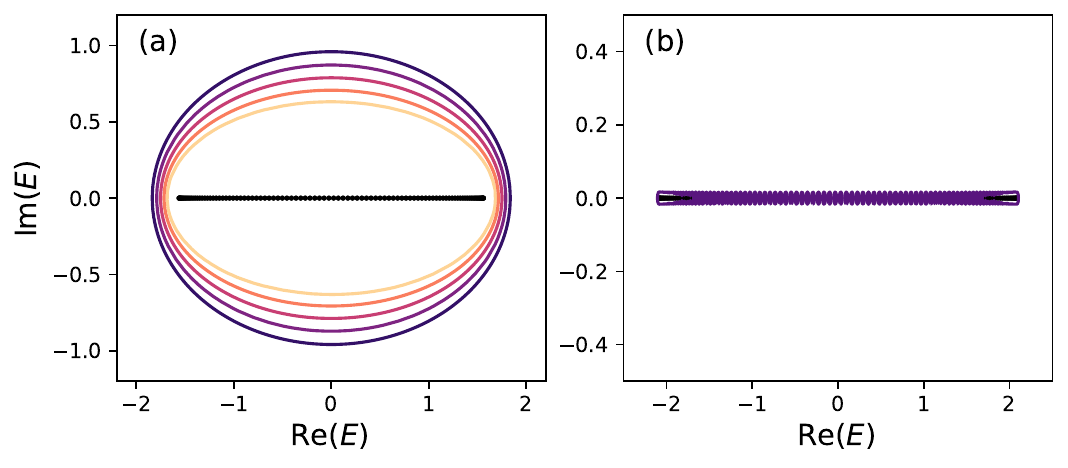}}
\caption{Boundaries of pseudospectra $\sigma_{\varepsilon}(H_{\text{SHN}})$. Open boundary condition. $\delta=0.5$, $L=100$ (a) $\gamma=0.8$. From the lightest to the darkest color, $\epsilon=10^{-15}\sim10^{-7}$. (b) $\gamma=0.4$.  From the lightest to the darkest color, $\epsilon=10^{-4}\sim10^{-2}$. The darkest purple color represents $10^{-2}$.  
In (a) and (b), the black solid points denote the exact energy eigenvalues under OBC. The pseudospectra is the interior enclosed by the boundary.}
\label{pseudospectraSyHN}
\end{figure}

\subsection{Symplectic Hatano-Nelson model}
In the previous subsection \ref{HNmodel}, we investigated the Hatano-Nelson model under OBC, whose condition number grows exponentially with the system size. Here, we consider a generalized version, the symplectic Hatano-Nelson model, in which the condition number transitions from exponential scaling with system size to a constant.
The model is defined as \cite{WindingSkin2,symplecticHN,NHSEEPTKawabata,pseudospectraNHSE2}: 
\begin{equation}
\begin{aligned}
H_{\text{SHN}}&=\sum_{j}\underline{c}^{\dagger}_{j+1}(J+\gamma\sigma_z-\mathrm{i}\delta\sigma_x)\underline{c}_{j} \\
&+\underline{c}^{\dagger}_{j}(J-\gamma\sigma_z+\mathrm{i}\delta\sigma_x)\underline{c}_{j+1}, 
\end{aligned}
\end{equation}
where $\underline{c}_j=(c_{j,A}, c_{j,B})^T$, with $c_{j,A}$ and $c_{j,B}$ representing annihilation operators of spinless fermions at site $j$ in chains $A$ and $B$. 
We define $T_{R}=J+\gamma\sigma_z-\mathrm{i}\delta\sigma_x$ and $T_{L}=J-\gamma\sigma_z+\mathrm{i}\delta\sigma_x$. Below we assume $J$, $\gamma$, and $\delta$ are non-negative real numbers and set $J=1$ as energy unit. 
Firstly, the symplectic Hatano-Nelson model satisfies the transpose version of time-reversal symmetry ($\text{TRS}^{\dagger}$) \cite{NHClassification,LiuChunHui1,LiuChunHui2}, given by $TH_{\text{SHN}}^{T}(k)T^{-1}=H(-k)$, where $T=\mathrm{i}\sigma_y$ is a unitary matrix. In real space, this symmetry is expressed as $T^{\prime}H_{\text{SHN}}^{T}{T^{\prime}}^{-1}=H$, in which $T^{\prime}=\Bbb{I}_{L\times L}\otimes \mathrm{i}\sigma_y$. Therefore, for every right (left) localized skin mode $|\psi_j^{R}\rangle$, there exists a corresponding oppositely localized mode $T^{\prime}|\psi_j^L\rangle^{*}$, which predicts both the degeneracy of the energy spectrum and the possibility of bidirectional localization. 
Such bidirectional localization, protected by a $\Bbb{Z}_2$ topological invariant, is dubbed as $\Bbb{Z}_{2}$ skin effect \cite{NHClassification,LiuChunHui1,LiuChunHui2,WindingSkin2}.

Under PBC, applying a Fourier transformation gives the eigenenergy  $E(k)=2J\text{cos}k\pm2\sqrt{\delta^2-\gamma^2}\text{sin}k$. If $\gamma>\delta$, the PBC spectrum holds the point gap in the complex plane, indicating the presence of NHSE. Conversely, when $\gamma<\delta$, the PBC energy spectrum collapses into the real line, signifying the absence of the NHSE \cite{WindingSkin,WindingSkin2}. 
Moreover, we find $[T_R, T_L]=0$, meaning that $T_R$ and $T_L$ can be diagonalized simultaneously. Specifically, we find $G^{-1}T_LG=\Lambda_L, G^{-1}T_RG=\Lambda_R$, in which $G=I-\dfrac{\delta}{\gamma+\sqrt{\gamma^2-\delta^2}}\sigma_y$, $\Lambda_L=\text{diag}\{J-\sqrt{\gamma^2-\delta^2}, J+\sqrt{\gamma^2-\delta^2}\}$, and $\Lambda_R=\text{diag}\{J+\sqrt{\gamma^2-\delta^2}, J-\sqrt{\gamma^2-\delta^2}\}$.
Applying the similarity transformation $H^{\prime}_{\text{SHN}}=(\Bbb{I}_{L\times L}\otimes G^{-1})H_{\text{SHN}}(\Bbb{I}_{L\times L}\otimes G)$, the Hamiltonian takes the form 
\begin{equation}
\begin{aligned}
H_{\text{SHN}}^{\prime}&=\sum_{j}\underline{c}_{j+1}^{\dagger}\Lambda_R\underline{c}_j+\underline{c}_j^{\dagger}\Lambda_L\underline{c}_{j+1} \\
&=\sum_{j}J_{+}c_{j+1,A}^{\dagger}c_{j,A}+J_{-}c^{\dagger}_{j,A}c_{j+1,A} \\ 
&+J_{-}c^{\dagger}_{j+1,B}c_{j,B}+J_{+}c_{j,B}^{\dagger}c_{j+1,B}, 
\end{aligned}
\end{equation}
in which $J_{+}=J+\sqrt{\gamma^2-\delta^2}$ and $J_{-}=J-\sqrt{\gamma^2-\delta^2}$.  For $\gamma>\delta$, the transformed Hamiltonian$H_{\text{SHN}}^{\prime}$ reduces to two decoupled Hatano-Nelson chains. Since the similarity transformation preserves the spectrum, the OBC eigenenergy is  $E^{(m)}_{\text{OBC}}=2\sqrt{J^2-\gamma^2+\delta^2}\text{cos}\theta_m$, $\theta_m=\dfrac{m\pi}{L+1}$, $m=\pm1, \pm2, ...\pm L$. For $\gamma<\delta$, the eigenenergy follows the same expression, but the model becomes Hermitian since $J_{+}={J_{-}}^*$.  
Although the Hamiltonian remains non-normal for any $\gamma\neq0$ and $\delta$, the condition number under OBC exhibits distinct scaling behaviors in the regimes $\gamma>\delta$ and $\gamma<\delta$. For $\gamma>\delta$, the condition number grows exponentially with system size: 
\begin{equation}
\begin{aligned}
\text{cond}(V)&=\text{cond}(G\underline{r}\oplus G\underline{r}^2\oplus\dots\oplus G\underline{r}^L) \\
&=\sqrt{\dfrac{R_{(L)}+\sqrt{R_{(L)}^2-4(1-(\dfrac{\delta}{\gamma})^2)}}{R_{(L)}-\sqrt{R_{(L)}^2-4(1-(\dfrac{\delta}{\gamma})^2)}}} \sim {r^{\prime}}^{L-1},
\end{aligned}
\end{equation}
in which $\underline{r}=\text{diag}\{\sqrt{J_{+}/J_{-}},  \sqrt{J_{-}/J_{+}}\}$, $ R_{(L)}={r^{\prime}}^{2L}+1/{r^{\prime}}^{2L}$, and $r^{\prime}=\sqrt{J_{+}/J_{-}}$. While for $\gamma<\delta$, the condition number remains bounded and independent of system size: 
\begin{equation}
\begin{aligned}
\text{cond}(V)=\text{cond}(\Bbb{I}_{L\times L}\otimes G) =\sqrt{\dfrac{\delta+\gamma}{\delta-\gamma}}.
\end{aligned}
\end{equation}

We emphasize again that although the value of the condition number depends on the choice of eigenvectors' gauge transformations, the qualitative distinction between exponential growth ($\gamma>\delta$) and size-independent behavior ($\gamma<\delta$) remains unchanged \cite{pseudospectraNHSE2}.
Correspondingly, the pseudospectra will display different characteristics in these two regimes. In the skin-effect regime ($\gamma>\delta$), perturbations can be exponentially amplified, rendering numerical results highly sensitive to round-off errors. As shown in Fig. \ref{pseudospectraSyHN}(a), the pseudospectra in this regime resemble those of the Hatano-Nelson model, indicating that default double precision may be insufficient for reliable diagonalization. Conversely, in the no-skin effect regime ($\gamma<\delta$), the spectrum remains stable, as illustrated in \ref{pseudospectraSyHN}(b). The pseudospectra behavior here is similar to normal matrices, ensuring that numerical calculations remain reliable with standard double precision. As for the PBC,  although $H_{\text{SHN}}$ is still non-normal,  the condition number is always finite and independent of $L$ \cite{pseudospectraNHSE2}.

\begin{figure}[htb]
\centering
\subfigure{
\includegraphics[height=8.0cm,width=8.4cm]{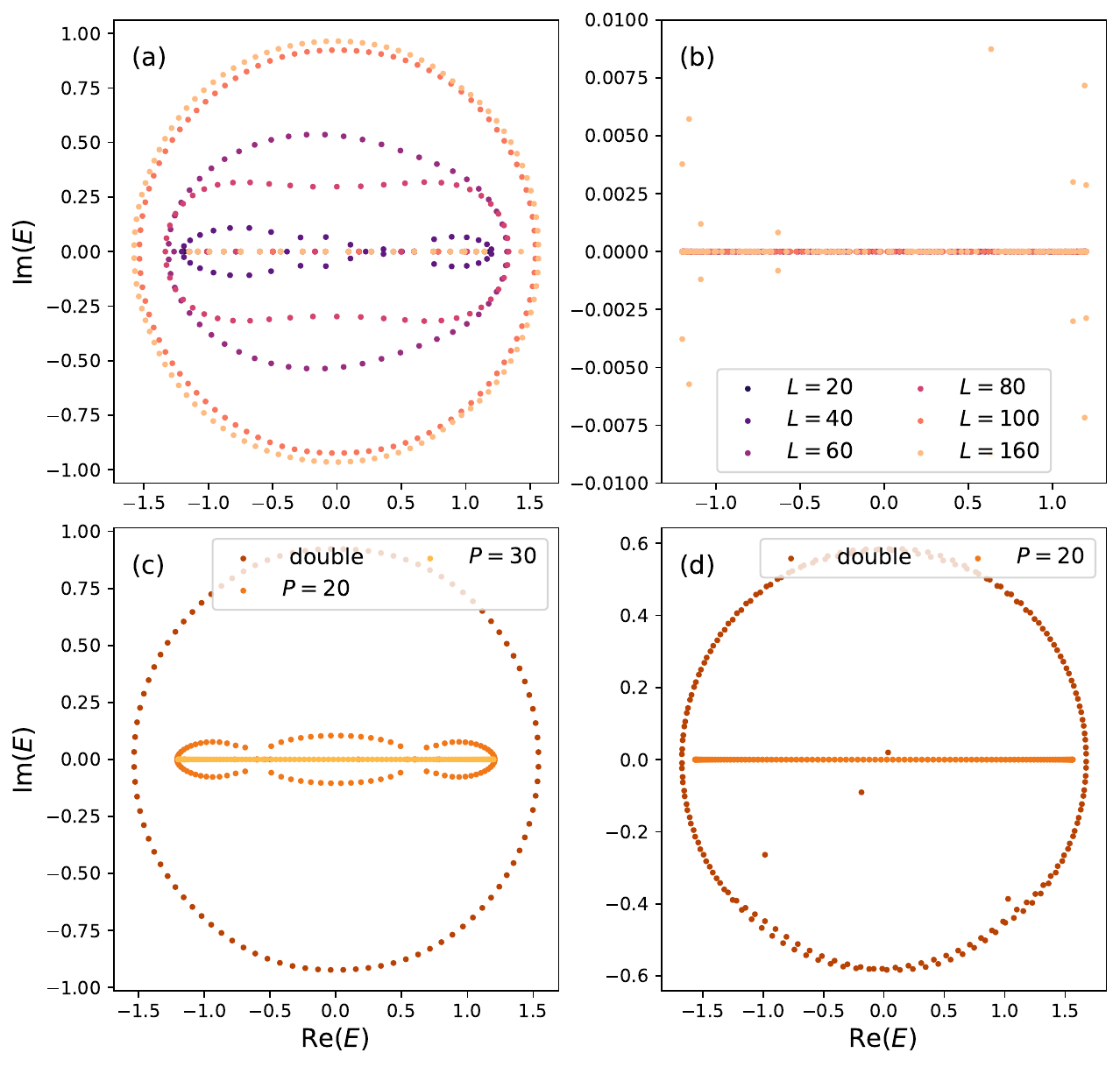}}
\caption{The numerical diagonalized energy spectrum of the Hatano-Nelson ((a), (b), (c)) and symplectic Hatano-Nelson model ((d)) under OBC. (a) $J=1.0$, $\gamma=0.8$. (b) $J=1.0$, $\gamma=-0.8$. (a) and (b) share the same legend. 
(c) $L=100$,  $J=1.0$, $\gamma=0.8$, $P$ represents the number of significant digits in precision. (d) The numerical results of the symplectic Hatano-Nelson model. $L=100$, $\delta=0.5$, $\gamma=0.8$.}
\label{NumericalDiag}
\end{figure}

\begin{figure*}[htb]
\centering
\subfigure{
\includegraphics[height=6.56cm,width=17.5cm]{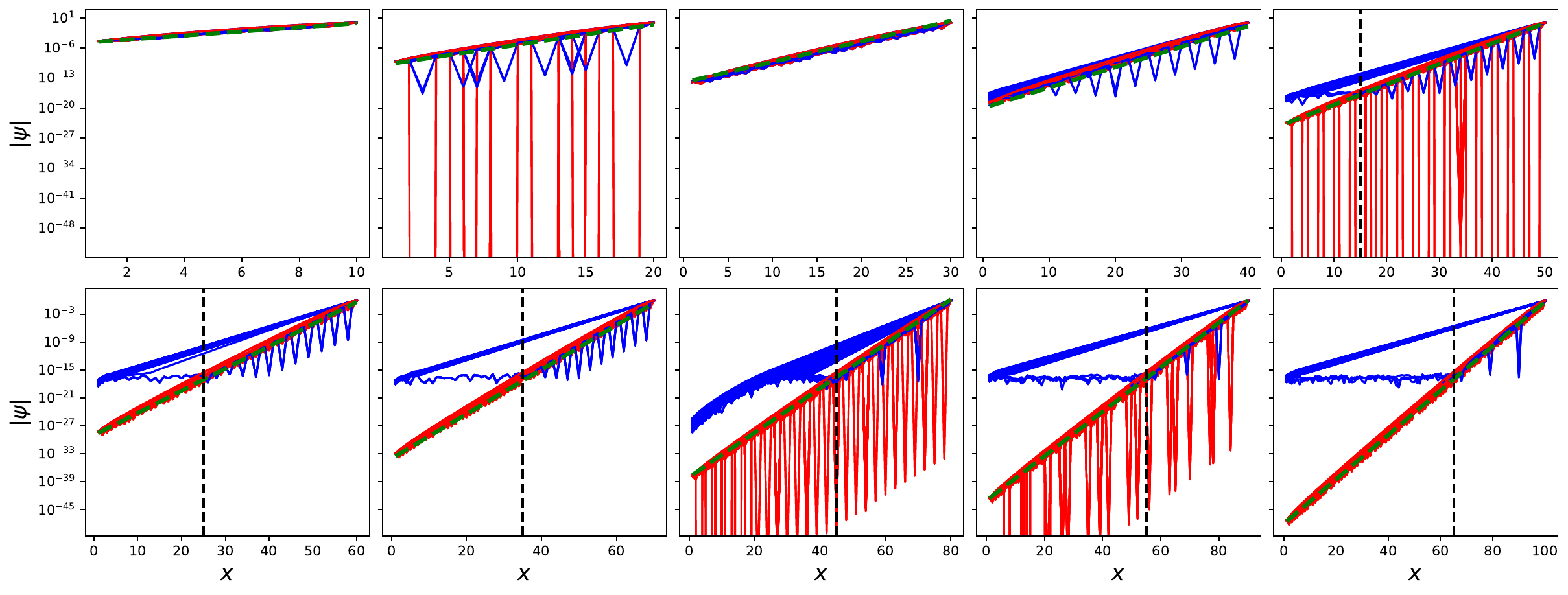}}
\caption{The numerical diagonalized wave functions of the Hatano-Nelson with different system sizes. The figure includes all wave functions for $\gamma=0.8$. The blue lines are the numerical results of double precision, while the red lines are the numerical results of the digit 50. The dashed green lines characterize the theoretical exponential localization of the Hatano-Nelson model with localization length $\xi=1/\ln\sqrt{(J+\gamma)/(J-\gamma)}$. Its form is $|\psi(x)|=ce^{x/\xi}$, $c$ varies with system size. It should be noted that the green lines are not the exact results of wave functions; they only approximately capture the exponential localization behaviors of the wave functions.
When $L\geq40$, the majority of wave functions with double precision display a larger localization length than the theoretical value, while the localization lengths with digit 50 closely match the theoretical value. We also observe that for double precision,  some eigenstates still obey the theoretical localization length. However, owing to the restrictions of the double precision, once the absolute value of the wave function decreases to around $10^{-16}\sim 10^{-15}$, it will get smooth.}
\label{NumericalDiagEigenVctors}
\end{figure*}

\section{Numerical errors of diagonalization}\label{SecNumericalErrorDiag}
In the previous section \ref{SecExamples}, we illustrate the spectral instability of the non-normal matrices utilizing the pseudospectra. In this section, we examine the numerical errors in the diagonalization of a non-Hermitian matrix under default double precision and specified digit precision. This analysis provides the foundation for understanding numerical errors in the dynamics of non-Hermitian Hamiltonians. 

Due to the spectral instability, the diagonalization results of highly non-normal matrices can be extremely sensitive to numerical precision.  
As shown in Fig. \ref{NumericalDiag}(a), for the Hatano-Nelson model under OBC with fixed $\gamma=0.8$ and default double precision, the energy eigenvalues remain real only for $L=20$. While for $L=40\sim 100$, the energy eigenvalues become complex, suggesting the unreliability of the exact diagonalization of a highly non-normal matrix, whose condition number grows exponentially with system size. Increasing numerical precision mitigates these errors, 
as illustrated in Fig. \ref{NumericalDiag}(c), where numerical inaccuracies gradually decrease. This behavior aligns with the interpretation that higher precision reduces the norm of perturbations in the context of pseudospectra. Similarly, as shown in \ref{NumericalDiag}(d), the diagonalization results of the symplectic Hatano-Nelson model in the skin regime ($\gamma>\delta$) display similar instabilities.

Interestingly, the diagonalization results for $\gamma=-0.8$ (see Fig. \ref{NumericalDiag}(b)) are significantly more numerically stable than those for $\gamma=0.8$ (see Fig. \ref{NumericalDiag}(a)), despite both non-Hermitian matrices being identical after a transpose operation. This discrepancy stems from the numerical diagonalization algorithm. To illustrate this, we consider the standard eigenvalue computation method: the $QR$ algorithm. The $QR$ algorithm follows these steps: first, it performs the $QR$ decomposition $H=Q_0R_0$, and then multiplies the $QR$ factors in the reverse order to obtain $H_1=R_0Q_0$. The algorithm then iterates this factor-and-reverse process $H_1=Q_1R_1$, $H_2=R_1Q_1$, $H_2=Q_2R_2\dots$, which eventually leads to a quasi-upper-triangular matrix,  composed of $1\times1$ blocks on the diagonal for real eigenvalues and $2\times2$ blocks for complex conjugate pairs of eigenvalues \cite{golubmatrix}. In Appendix \ref{AppendixNumericalAsymmetry}, we simulate the diagonalization routine and demonstrate that the numerical asymmetry originates from the $QR$ algorithm. Specifically, we find that the matrices $H_1, H_2, H_3, \dots$ generated during the $QR$ iterations exhibit distinct structural differences between $\gamma = 0.8$ and $\gamma = -0.8$. For $\gamma = 0.8$, the $QR$ decomposition will produce extremely small denominators during the Householder reflections, leading to numerical instability. In contrast, for $\gamma = -0.8$, such small denominators are avoided, resulting in improved numerical stability. 
Furthermore, as shown in Fig. \ref{WavefunctionMinusGam}, we observe that the numerical results of wave functions for $\gamma=-0.8$ can be far more accurate than $10^{-16}$, even with double precision. The exact reason for this remarkable accuracy in the case of negative $\gamma$ remains elusive and warrants further investigation. We emphasize that for large system size $L$, such as $L=160$ in Fig. \ref{NumericalDiag}(b), the energy eigenvalues will also become complex, i.e., incorrect, which means the numerical instability still exists for $\gamma=-0.8$.  

Besides the energy spectrum, we numerically calculate the OBC wave functions of the Hatano-Nelson model. Theoretically, the OBC wave function can be acquired  analytically through non-Bloch band theory \cite{non-bloch,GBZ}
\begin{equation}
|E^{(m)}_{\text{OBC}}\rangle\propto \left( 
\begin{matrix}
r^1 \sin \left(\theta_m \right) \\
r^2 \sin \left( 2\theta_m\right) \\
\vdots \\
r^{L-1} \sin \left( (L-1)\theta_m \right) \\
r^L \sin \left( L\theta_m \right)
\end{matrix} \right),
\end{equation}
which describes the exponential decay of wave function with localization length $\xi=1/\ln(r)=1/\ln\sqrt{(J+\gamma)/(J-\gamma)}$.
Under the normalization condition $\Vert|E^{(m)}_{\text{OBC}}\rangle\rVert^2=1$, which is automatically enforced by most numerical diagonalization algorithms, the absolute values of wave function components near the boundary can become much smaller than $10^{-16}$ for large $L$. This precision limit implies that eigenstates obtained with double precision become unreliable. Indeed, as shown in Fig. \ref{NumericalDiagEigenVctors}, when $L\gtrsim40$, the numerically obtained wave functions deviate from theoretical predictions. These incorrect results can be categorized into two types: 1) Some wave functions decay exponentially with the expected rate $1/\xi$ until their amplitudes reach approximately $10^{-16}$, after which they remain at that level rather than continuing to decay (see also a similar pattern in spectral decomposition of variational quantum states \cite{ET2024}). 2) Other wave functions decay exponentially but at a rate smaller than the predicted $1/\xi$, ensuring that the smallest wave function amplitude at the left boundary remains close to $10^{-16}$, the approximate lower bound of double precision. When numerical precision is increased to 50-digit precision, this lower bound improves to at least $10^{-50}$, significantly smaller than the absolute values of the wave function at all sites. Consequently, as displayed in Fig. \ref{NumericalDiagEigenVctors}, the wave functions obtained with 50-digit precision roughly agree with analytical results.
Moreover, we also improve the precision from double to digit $20$ and digit $30$. As shown in Fig. \ref{ConnectionSupp1} and \ref{ConnectionSupp2}, the diagonalized wave functions get closer and closer to the analytical results. Interestingly, we will demonstrate in Sec. \ref{SecNHEvo} that the incorrectness of the diagonalized wave functions is closely related to the incorrectness of the non-Hermitian evolution. Even though this incorrectness of the wave functions is extremely small, the incorrectness of the corresponding non-Hermitian evolution can be significant.

\begin{figure*}[htb]
\centering
\subfigure{
\includegraphics[height=9.9cm,width=16.5cm]{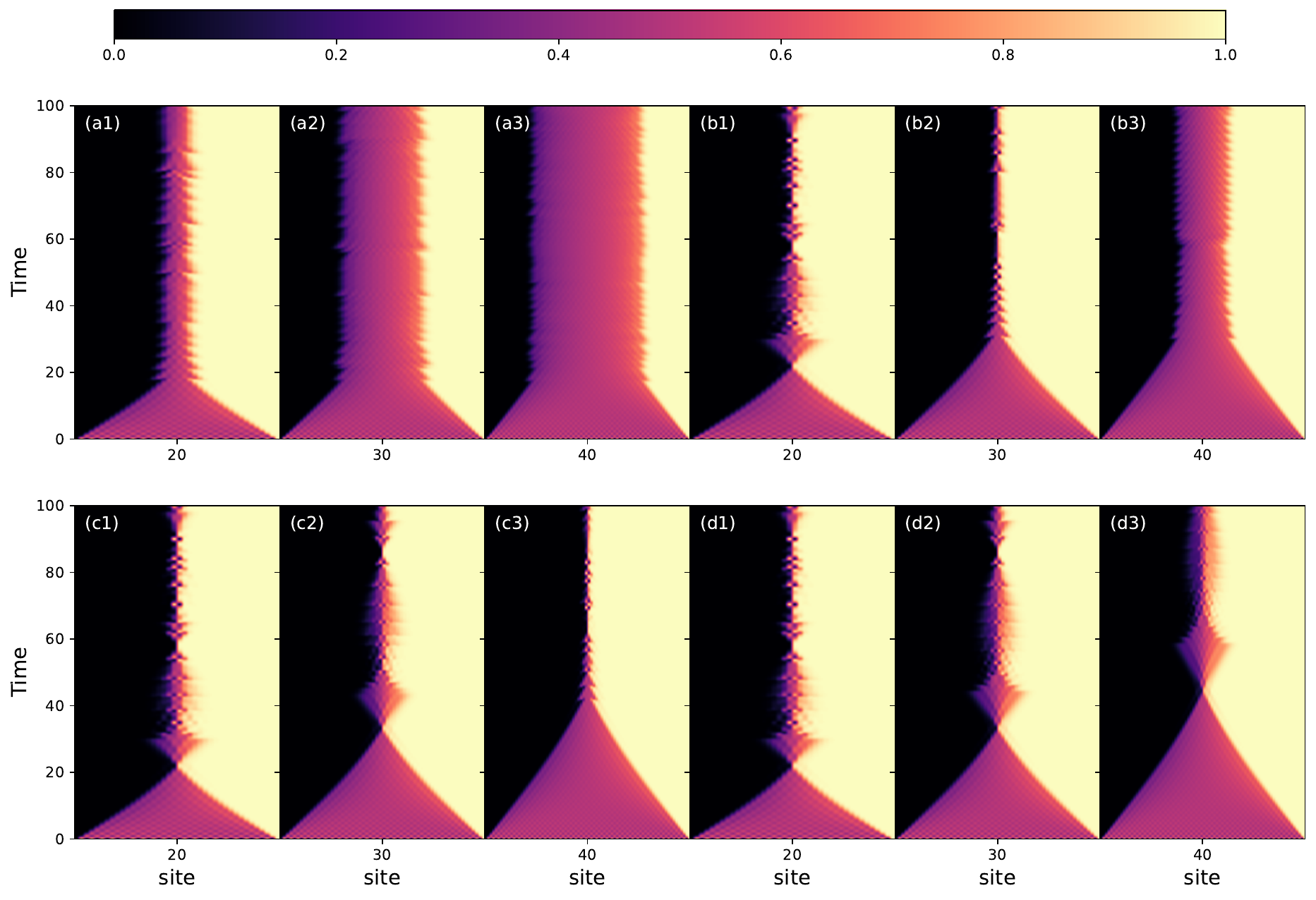}}
\caption{The density evolution of the Hatano-Nelson model under OBC with different system sizes and numerical precisions. The initial state is N\'{e}el state $|0101\cdots01\rangle$. $\gamma=0.8$. Each subfigure is annotated with half the system size $L/2$.
(a) Double precision. (b) Digit $P=20$. (c) Digit $P=30$. (d) Digit $P=50$.}
\label{nonHermitainEvo}
\end{figure*} 

\begin{figure}[htb]
\centering
\subfigure{
\includegraphics[height=4.8cm,width=8.4cm]{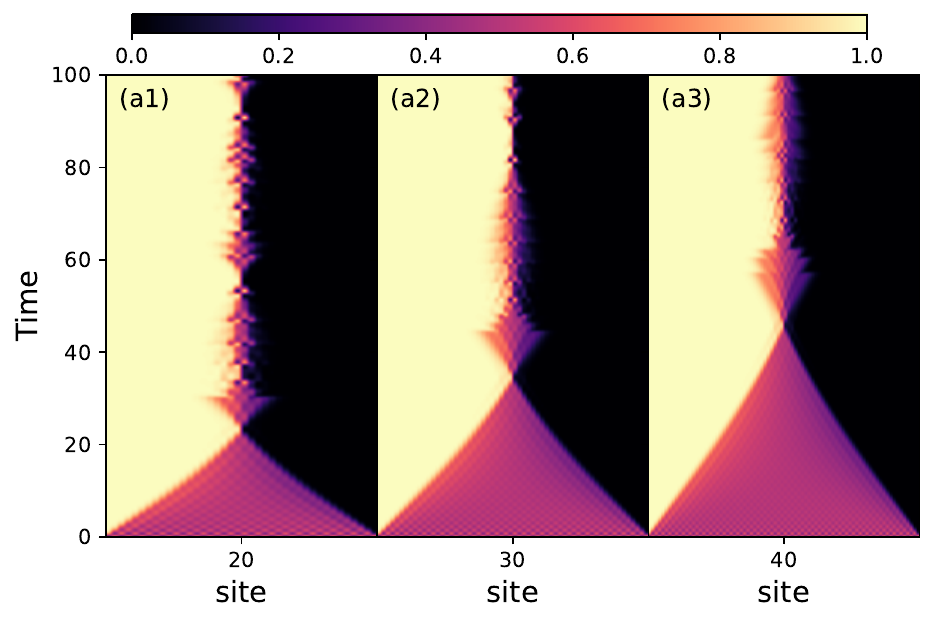}}
\caption{The density evolution of the Hatano-Nelson model under OBC with double precision. The initial state is N\'{e}el state $|0101...01\rangle$. $\gamma=-0.8$. Each subfigure is annotated with half the system size $L/2$.}
\label{DensityEvoMinusGam}
\end{figure}

\begin{figure*}[htb]
\centering
\subfigure{
\includegraphics[height=9.6cm,width=16.0cm]{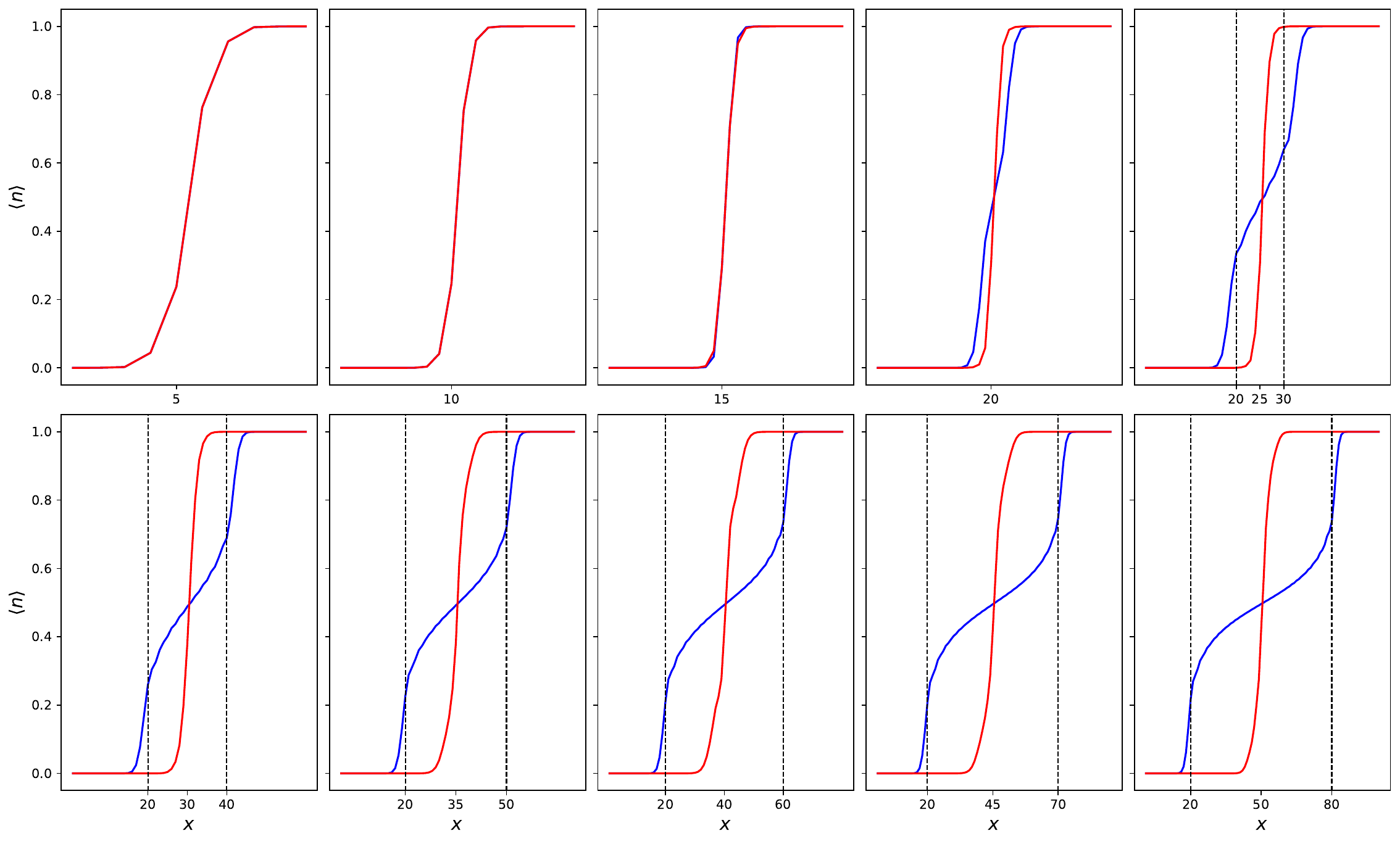}}
\caption{The steady-state density distribution of the Hatano-Nelson model under OBC with various system sizes ($L=10\sim100$). Each subfigure is annotated with half the system size. The numerical precisions adopted are default double precision (blue) and digit 50 (red). The width of the relatively smooth region is about $L-40$ (black dashed lines). $\gamma=0.8$.}
\label{density}
\end{figure*}

In terms of the symplectic Hatano-Nelson model, the exact form of the wave functions can also be derived.
Using the analytical solution of the Hatano-Nelson model, the two degenerate OBC wave functions with eigenenergy ${E^{\prime}}^{(m)}$ of $H^{\prime}_{\text{SHN}}$ (i.e., the symplectic Hatano-Nelson model after the similarity transformation) can be written as 
\begin{equation}
|{E^{\prime}}^{(m)}_{\text{OBC}}\rangle\propto \left( 
\begin{matrix}
{r^{\prime}} \sin \left( \theta_m \right) \\ 0 \\
{r^{\prime}}^2 \sin \left( 2\theta_m \right) \\ 0 \\
\vdots \\
{r^{\prime}}^{L} \sin \left( L\theta_m\right) \\
0 
\end{matrix} \right),
\left( 
\begin{matrix}
0 \\ {r^{\prime}}^{-1}\text{sin}(\theta_m) \\
0\\ {r^{\prime}}^{-2} \text{sin}(2\theta_m) \\
\vdots \\
0 \\
{r^{\prime}}^{-L} \sin \left(L\theta_m \right)
\end{matrix} \right), 
\end{equation}
in which $r^{\prime}=\sqrt{J_+/J_-}=\sqrt{\dfrac{J+\sqrt{\gamma^2-\delta^2}}{J-\sqrt{\gamma^2-\delta^2}}}$. Obviously, if $\gamma<\delta$, $|r^{\prime}|=1$, which reconfirms the condition of no skin effect.
Since the Hamiltonian satisfies $H_{\text{SHN}}=(\Bbb{I}_{L\times L}\otimes G)H_{\text{SHN}}^{\prime}(\Bbb{I}_{L\times L}\otimes G^{-1})$, the two degenerate eigenvectors of the symplectic Hatano-Nelson model take the form
\begin{equation}
|E^{(m)}\rangle_{\text{OBC}}\propto \left( 
\begin{matrix}
{r^{\prime}}^1 \sin \left( \theta_m \right) \\c^{\prime}{r^{\prime}}^1\text{sin}(\theta_m) \\
{r^{\prime}}^2 \sin \left( 2\theta_m \right) \\ c^{\prime}{r^{\prime}}^2\text{sin}(2\theta)_m \\
\vdots \\
{r^{\prime}}^{L} \sin \left( L\theta_m\right) \\
c^{\prime}{r^{\prime}}^L\text{sin}(L\theta_m)
\end{matrix} \right),
\left( 
\begin{matrix}
-c^{\prime}{r^{\prime}}^{-1}\text{sin}(\theta_m) \\ {r^{\prime}}^{-1}\text{sin}(\theta_m) \\
-c^{\prime}{r^{\prime}}^{-2}\text{sin}(2\theta_m) \\ {r^{\prime}}^{-2} \text{sin}(2\theta_m) \\
\vdots \\
-c^{\prime}{r^{\prime}}^{-L}\text{sin}(L\theta_m) \\
{r^{\prime}}^{-L} \sin \left(L\theta_m \right)
\end{matrix} \right),
\end{equation}
in which $c^{\prime}=G_{2,1}=-\mathrm{i}\delta/(\gamma+\sqrt{\gamma^2-\delta^2})$. Clearly, for a given eigenenergy $E^{(m)}_\text{OBC}$, both right- and left-localized eigenstates exist, reflecting $\Bbb{Z}_{2}$ skin effect.
Now, it is apparent that the wave functions of the symplectic Hatano-Nelson model in the skin regime ($\gamma>\delta$) also suffer similar difficulties with the Hatano-Nelson model when performing numerical diagonalization. 
Enforcing the normalization condition $\lVert |{E^{\prime}}^{(m)}_{\text{OBC}}\rangle\rVert^2=1$, the theoretical values of the wave functions around the boundary can be much smaller than the minimum precision due to the exponential localization behavior of the wave functions, leading to numerical errors. 
These numerical errors can be reduced by improving the numerical precision, specifically by lowering the minimum precision threshold.
For practical numerical diagonalization, the computed wave functions of the symplectic Hatano-Nelson model are often superpositions of right- and left-localized eigenstates, which is less illustrative than the Hatano-Nelson model and not presented here.

\section{Numerical errors of non-Hermitian evolution}
\label{SecNHEvo}
In this section, we examine the non-Hermitian Hamiltonian evolution: 
\begin{equation}
|\psi(t)\rangle=\dfrac{e^{-\mathrm{i}Ht}|\psi_0\rangle}{\lVert e^{-\mathrm{i}Ht}|\psi_0\rangle \rVert}
\end{equation}
The two models introduced in section \ref{SecExamples} are $U(1)$-symmetric and non-interacting, ensuring the preservation of Gaussianity. Their non-Hermitian evolution has been previously simulated using Gaussian-state methods \cite{NHSEEPTKawabata}. 
In this study, we also employ the Gaussian-state simulation method from Ref. \cite{NHSEEPTKawabata} for comparison. 
The observables of interest—correlation, density distribution, particle current, and entanglement entropy can be calculated using the Gaussian state simulation method. A brief overview of the simulation method and the definitions of these observables are provided in Appendix \ref{AppendixGaussianStateSimulation}.

\begin{figure}[htb]
\centering
\subfigure{
\includegraphics[height=8.4cm,width=8.4cm]{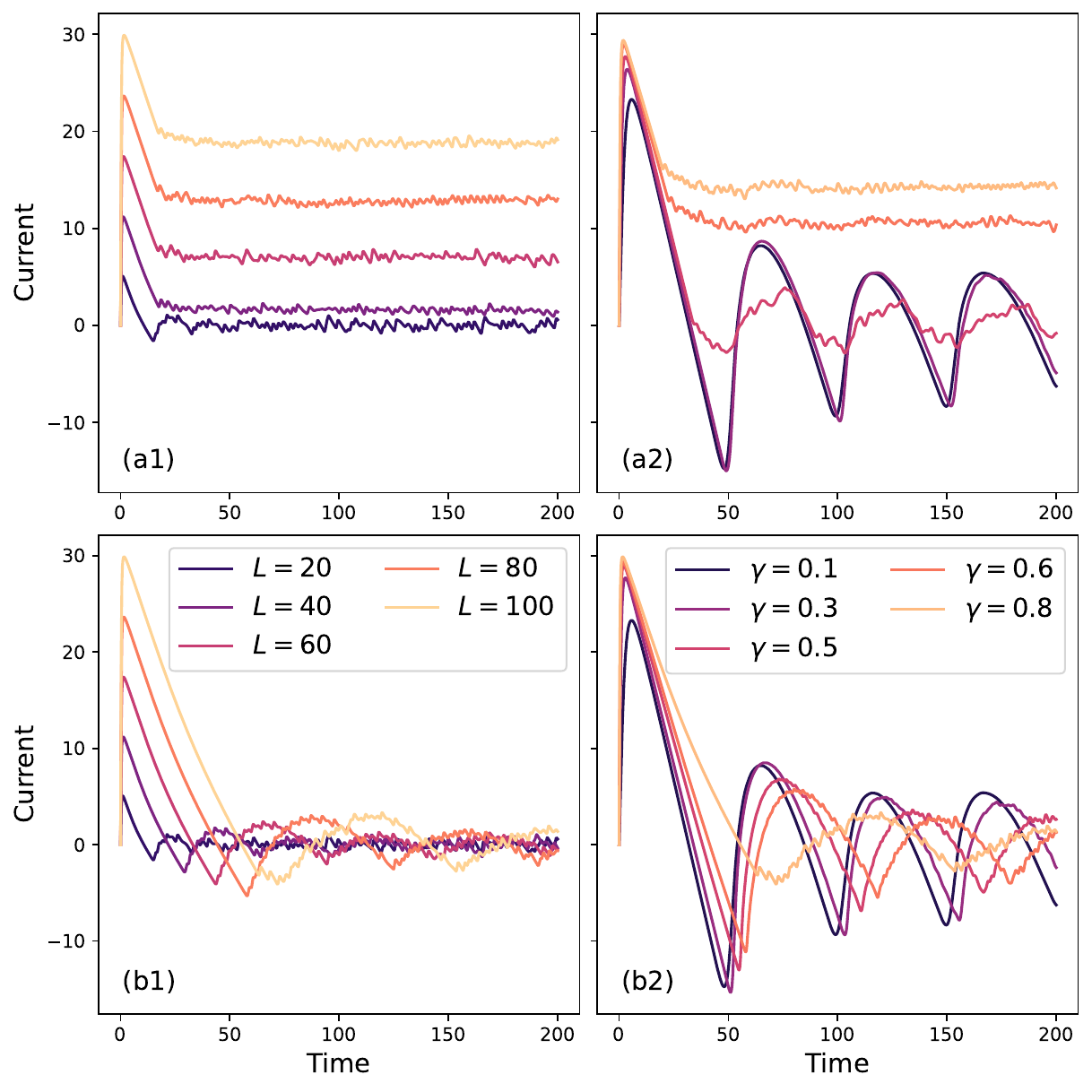}}
\caption{The evolution of the current of the Hatano-Nelson model under OBC. The upper panels ((a1), (a2)) are the results of double precision. The lower panels ((b1), (b2)) are the results of digit $50$. (a1) and (b1) share the same legend, and $\gamma=0.8$. (b1) and (b2) share the same legend, and  $L=100$. }
\label{CurrentEvoHN}
\end{figure}

\begin{figure}[htb]
\centering
\subfigure{
\includegraphics[height=6.0cm,width=8.4cm]{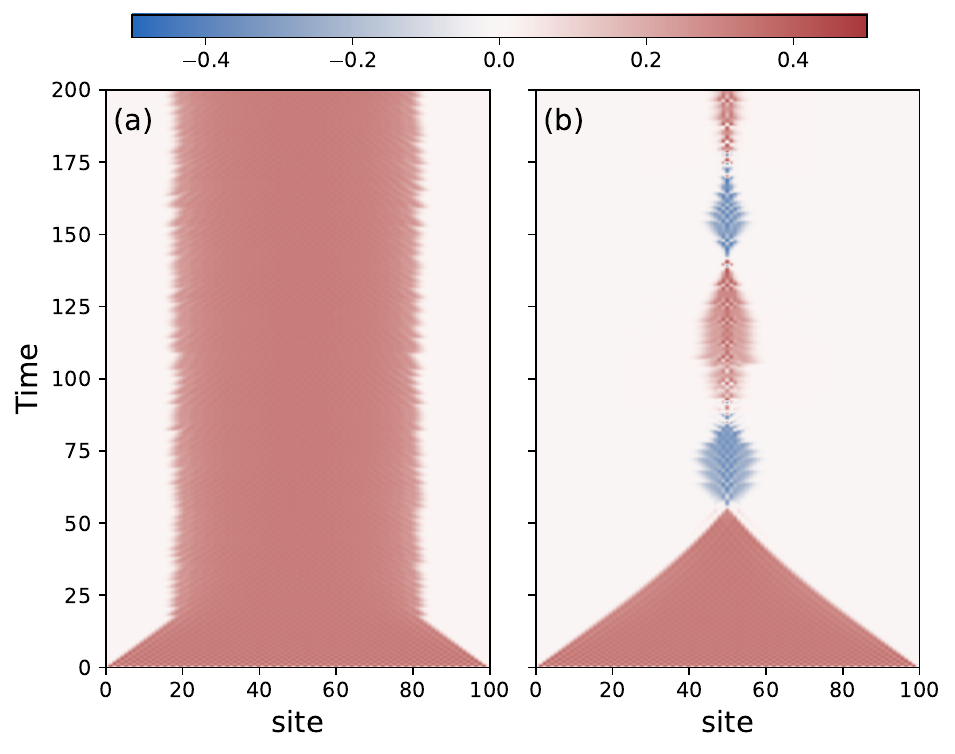}}
\caption{ The evolution of the local current of the Hatano-Nelson model under OBC. $\gamma=0.8$. (a) Double precision. (b) Digit $50$.}
\label{LocalCurrentEvo}
\end{figure}

We find that some key numerical results from Refs. \cite{NHSEEPTKawabata,Schiro} are \textbf{incorrect} due to the ignorance of numerical errors and the use of double precision, resulting in misleading physical pictures. For instance, Ref. \cite{NHSEEPTKawabata} incorrectly reports the emergence of a source and drain in the steady state of the Hatano-Nelson model, with a smooth transition region in the middle where particle occupation changes gradually and local current never vanishes. 

On the contrary, the correct steady state consists of approximately half of the system fully occupied and the other half completely unoccupied (half-filled system), without any transition region. As shown in Fig. \ref{nonHermitainEvo}, this incorrect transition region gradually disappears as numerical precision increases (see panels (a3), (b3), and (d3)). More specifically, for the steady-state density distribution with double precision in Fig. \ref{density}, we observe that the widths of the fully occupied and unoccupied regions remain nearly invariant with system size.  Interestingly, we find that the width of the incorrect middle region in non-Hermitian evolution (see Fig. \ref{density}) roughly matches the width of the incorrect regime in eigenvector computation (see Fig. \ref{NumericalDiagEigenVctors}). For example, at double precision, the width of the incorrect region in eigenvector calculations is approximately $L-35$ (see Fig. \ref{NumericalDiagEigenVctors}, black dashed lines), while the width of the incorrect middle region in non-Hermitian evolution is about $L-40$ (see Fig. \ref{density}, dashed lines).
This relationship holds for other numerical precisions as well (digit $20$, $30$, see Fig. \ref{ConnectionSupp1} and \ref{ConnectionSupp2}), suggesting a connection between numerical errors in non-Hermitian evolution and those in eigenvector diagonalization, which we will discuss later.

Additionally, it is reasonable that the steady state computed with double precision exhibits a weaker skin effect than that obtained with higher precision. The skin effect arises from non-reciprocity, but numerical errors can be treated as random perturbations, partially reducing non-reciprocity and thereby weakening the skin effect. 
Since numerical errors (perturbation strength) decrease with increasing precision, the skin effect strengthens accordingly (see Fig. \ref{nonHermitainEvo}, panels (a3), (b3), and (d3)).
Remarkably, as previously shown, the numerically diagonalized energy spectrum and wave functions with $\gamma=-0.8$ are much more accurate than $\gamma=0.8$ at double precision. Likewise, Fig. \ref{DensityEvoMinusGam} also supports that the non-Hermitian evolution results of $\gamma=-0.8$ are numerically more accurate than $\gamma=0.8$. Specifically, with default double precision, Fig. \ref{DensityEvoMinusGam} (a1), (a2), (a3) already provides correct simulation results, while the corresponding results of $\gamma=0.8$ in Fig. \ref{nonHermitainEvo} (a1), (a2), (a3) are wrong (with middle region). 

Furthermore, Ref. \cite{NHSEEPTKawabata} claims that 
above a certain threshold for non-reciprocity $\gamma$ or system size $L$, a nonzero charge current emerges (see Fig. 4(a), 4(b), and 4(c) in Ref. \cite{NHSEEPTKawabata}). However, this conclusion is physically incorrect due to numerical errors.
In Ref. \cite{NHSEEPTKawabata}, the non-zero current only emerges when $\gamma$ or $L$ exceeds a certain threshold, coinciding with the onset of strong non-normality. We replicate the phenomenon in Fig. \ref{CurrentEvoHN}(a1) and (a2) using double precision, which confirms a stable late-time current for large $\gamma$ and $L$.
However, as depicted in Fig. \ref{CurrentEvoHN}(b1) and (b2), once the numerical precision is increased to digit $50$, the current consistently fluctuates around zero, and the late-time average of the charge current vanishes for all values of $\gamma$ and $L$. Thus, the results in Fig. 4(a), 4(b), and 4(c) of Ref. \cite{NHSEEPTKawabata} lack physical meaning. 
Regarding the local current, Ref. \cite{NHSEEPTKawabata} (see Fig. 5 in Ref. \cite{NHSEEPTKawabata}) presents results similar to our Fig. \ref{LocalCurrentEvo}(a). To explain the non-zero local current in the middle region, Ref. \cite{NHSEEPTKawabata} proposes that the particles are injected around the site $20$ from the environment and leave out of the system around the site $80$, thereby generating a non-zero current between the ``source" and ``drain".  
With the numerical precision digit $50$, the middle current flow region vanishes, as illustrated in Fig. \ref{LocalCurrentEvo}(b), which is intuitive. Because the non-reciprocity tends to drive the particles toward one side constantly, the steady state should be about the half-side occupied. As a result, particles in the steady state are almost frozen due to the Pauli exclusion principle, and the current should approach zero. 
In summary, the numerical errors in Ref. \cite{NHSEEPTKawabata} lead to the erroneous conclusion that non-Hermitian evolution supports a nonzero steady-state charge current.

\begin{figure}[htb]
\centering
\subfigure{
\includegraphics[height=4.2cm,width=8.4cm]{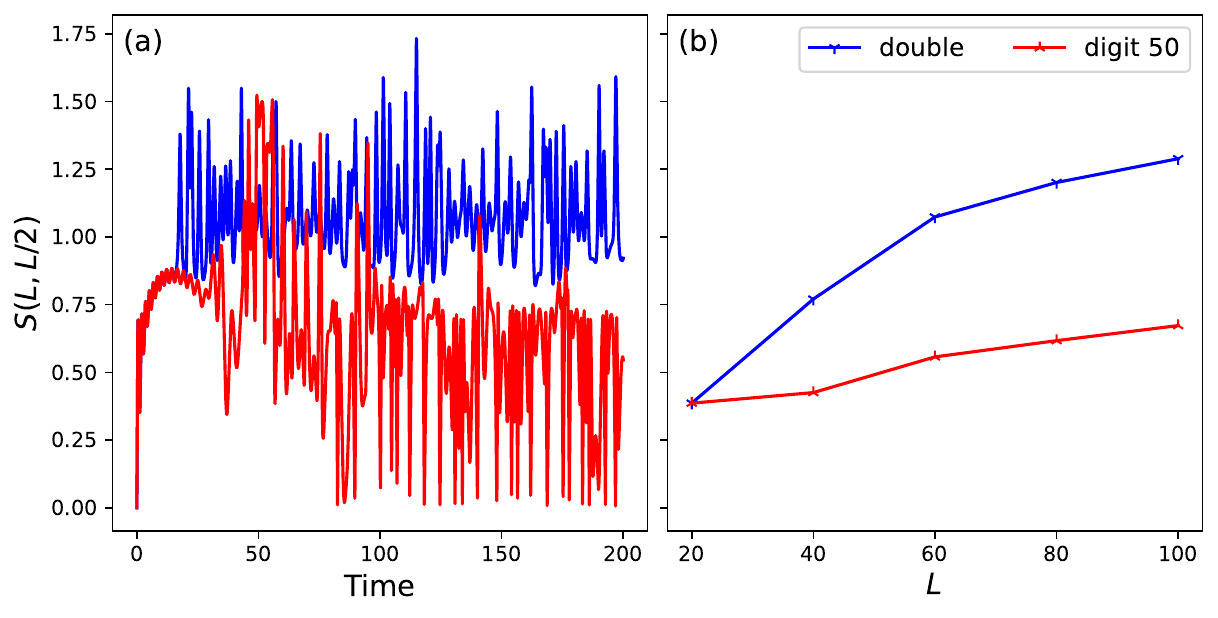}}
\caption{Entanglement dynamics and 
 steady-state entanglement entropy of the Hatano-Nelson model under OBC with double precision (blue) and digit $50$ (red). $L=60$, $\gamma=0.8$. (a)  The dynamics of entanglement entropy. (b) Steady-state entanglement entropy.}
\label{EEntropyHN}
\end{figure}

In terms of entanglement, because the numerically correct steady state tends to be more frozen---namely, the particles can scarcely move due to the Pauli exclusion principle---it is reasonable to predict that the correct entanglement should be smaller than the results presented in Ref. \cite{NHSEEPTKawabata}.
Indeed, as shown in Fig. \ref{EEntropyHN}, the entanglement entropy computed with 50-digit precision is consistently smaller than that obtained with double precision, except for $L=20$, where double precision is already sufficient for accurate results. Therefore, the qualitative conclusion from Ref. \cite{NHSEEPTKawabata} that the steady-state entanglement obeys an area law for $\gamma\neq0$ remains valid.

Similarly, Ref. \cite{Schiro} utilizes the Faber polynomial method to investigate the non-Hermitian evolution of the Hatano-Nelson model as well. However, the numerical results in Ref. \cite{Schiro} also reproduce the wrong smooth middle region and non-zero charge current, which implies that the Faber polynomial method cannot get rid of the numerical instability induced by strong normality. 

\begin{figure}[htb]
\centering
\subfigure{
\includegraphics[height=8.0cm,width=7.0cm]{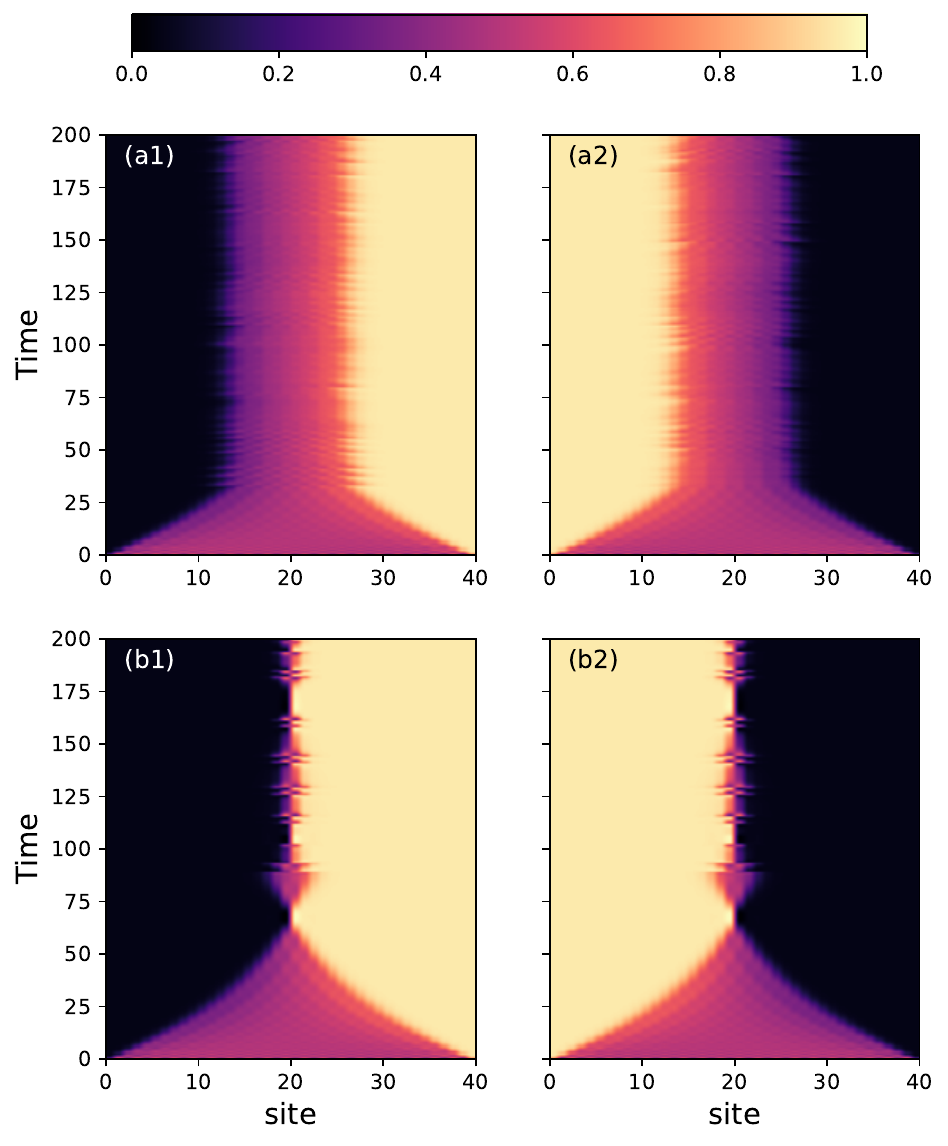}}
\caption{The density evolution of the symplectic Hatano-Nelson model under OBC. $L=40$, $\gamma=1.0$, $\delta=0.4$. The initial state is N\'eel state $|1010\cdots10\rangle_A|0101\cdots01\rangle_B$.
The results in (a1) and (a2) are simulated with double precision, while (b1) and (b2) are simulated with digit 30. (a1) and (b1) are the density evolution of the chain $A$, while (a2) and (b2) are the density evolution of chain $B$.}
\label{SymplecticHNEvo}
\end{figure}

\begin{figure}[htb]
\centering
\subfigure{
\includegraphics[height=4.2cm,width=8.4cm]{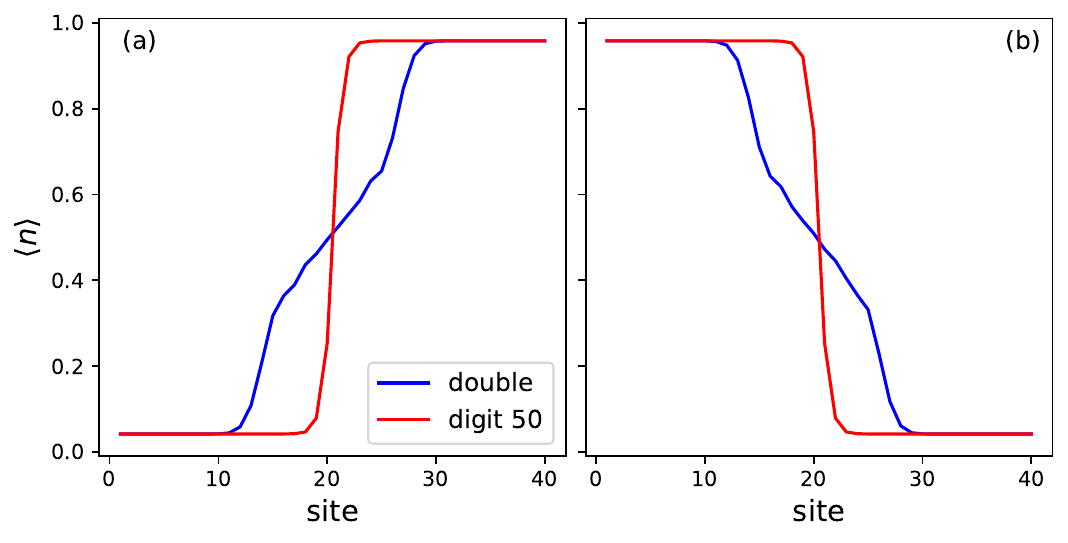}}
\caption{The steady-state density of the symplectic Hatano-Nelson model under OBC. $L=40$, $\gamma=1.0$, $\delta=0.4$. The results with the color blue are simulated with double precision, while the results with red are simulated with the digit $50$. (a) The steady-state density of the chain $A$. (b) The steady-state density of chain $B$. The densities around the two boundaries are about 0.9583 and 0.0417, respectively. }
\label{SymplecticHNSteady}
\end{figure}

\begin{figure}[htb]
\centering
\subfigure{
\includegraphics[height=4.2cm,width=8.4cm]{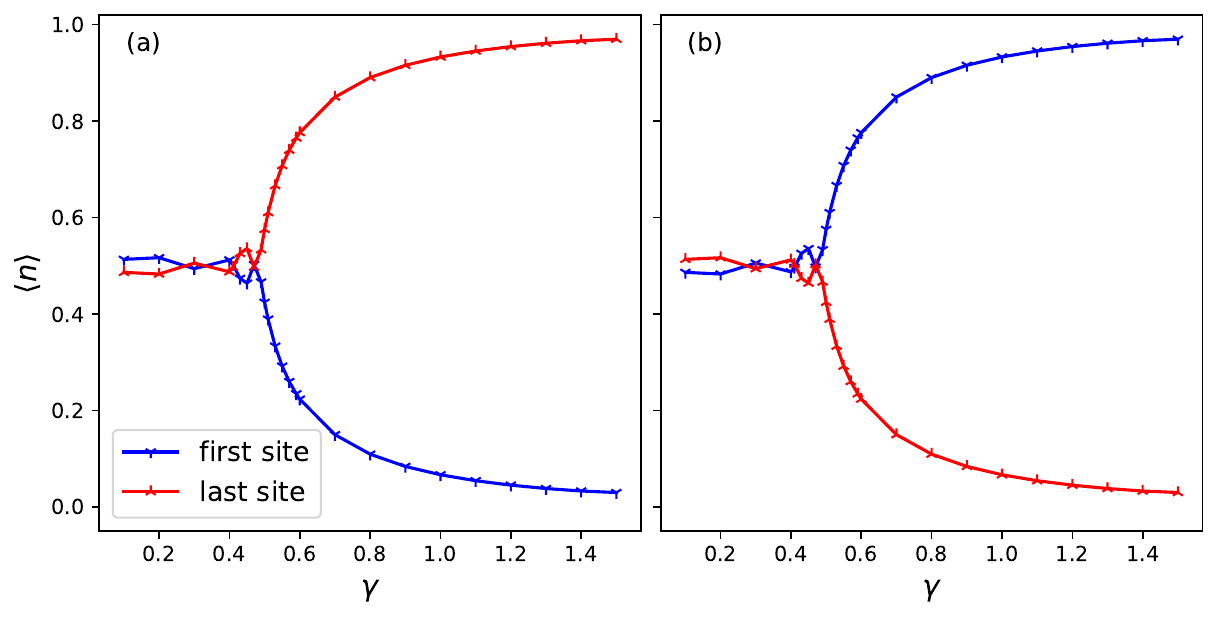}}
\caption{The steady-state edge density ($\langle n_{1,s}\rangle$, $\langle n_{L,s}\rangle$, $s=\{A,B\}$) of the symplectic Hatano-Nelson model under OBC. $\delta=0.5$. (a) The steady-state edge density of the chain $A$. (b) The steady-state density of chain $B$.}
\label{residuecharge}
\end{figure}

\begin{figure}[htb]
\centering
\subfigure{
\includegraphics[height=4.2cm,width=8.4cm]{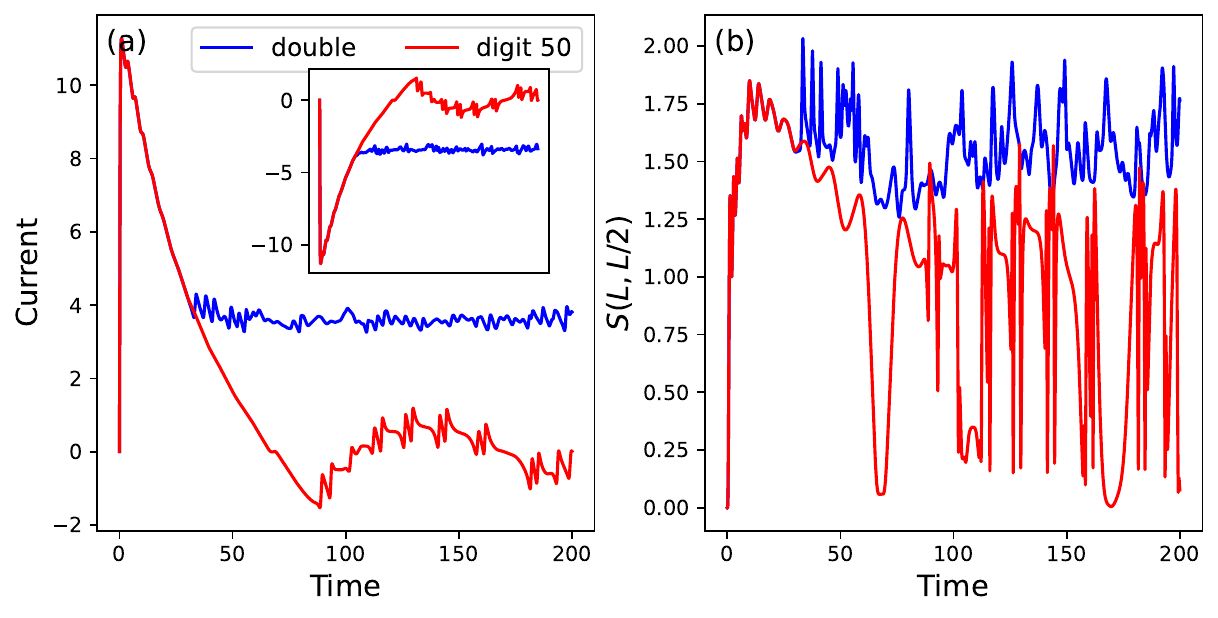}}
\caption{The total current and entanglement entropy of the symplectic Hatano-Nelson model in chain $A$ and chain $B$, respectively.  $L=40$, $\gamma=1.0$, $\delta=0.4$. (a) The main plot (inset) describes the total current in chain $A$ ($B$). $I_{A}+I_{B}=0$.  (b) Entanglement dynamics}
\label{EEntropySymplecticHN}
\end{figure}

We now investigate the non-Hermitian evolution of the symplectic Hatano-Nelson model. 
In the many-body Hatano-Nelson model, the steady state is intuitively expected to be half occupied and half unoccupied due to the unidirectional localization. However, in the symplectic Hatano-Nelson model, owing to bidirectional localization, the long-time steady state is less straightforward to predict. 

The numerical simulation results for the density are presented in Fig. \ref{SymplecticHNEvo} and Fig. \ref{SymplecticHNSteady}. First of all, the particles in chain $A$ ($B$) tend to localize on the right (left) side, although the average particle distribution across both chains is uniform due to reciprocity. However, as shown in Fig. \ref{SymplecticHNSteady} and \ref{residuecharge},  different from the Hatano-Nelson model, in which the system is completely occupied or unoccupied around the edges, there is residue density around the boundary, and the residue density satisfies $\langle n_1\rangle+\langle n_L\rangle=1$. Moreover, the numerical results show that the residue density is independent of the system size $L$ (not shown).
Specifically, as depicted in Fig. \ref{residuecharge},  for a fixed value of $\delta$, the residue charge $\langle n_{1,A}\rangle$ ($\langle n_{L,A}\rangle$) gradually approaches zero (one) with increasing $\gamma$. It is consistent with the previous results of the Hatano-Nelson model since in the limit of $\gamma/\delta\rightarrow\infty$, the symplectic Hatano-Nelson model can be regarded as two decoupled Hatano-Nelson models. As for $\gamma<\delta$, the residue charge is close to $0.5$, which corresponds to the no-skin effect scenario.  

Akin to the Hatano-Nelson model, the numerical results of the symplectic Hatano-Nelson model in Figs. 8, 9, 10, 11 of Ref. \cite{NHSEEPTKawabata} (including density, correlation, current,  entanglement) are also incorrect due to numerical errors. Firstly, as indicated in Fig. \ref{SymplecticHNEvo} and \ref{SymplecticHNSteady}, the middle smooth varying region disappears when the numerical precision is improved from double to digit $50$. Moreover, as shown in Fig. \ref{EEntropySymplecticHN}, the current in chain $A$ or $B$ with digit 50 (i.e. the spin current in Fig. 10(b), 10(c), 10(d) of Ref. \cite{NHSEEPTKawabata}) vanishes after long-time averaging, in contrast to the results with double precision, where a non-zero current exists in the late time. Furthermore, Fig. \ref{EEntropySymplecticHN}(b) shows that the steady-state entanglement at digit $50$ in the skin regime ($\gamma>\delta$) is smaller than the results with double precision as well. It is vital to note that we only present the results of $\gamma>\delta$ for the symplectic Hatano-Nelson model since the results of $\gamma<\delta$ are numerically stable owing to the disappearance of the skin effect, i.e., small condition number.

\begin{figure}[htb]
\centering
\subfigure{
\includegraphics[height=4.2cm,width=8.4cm]{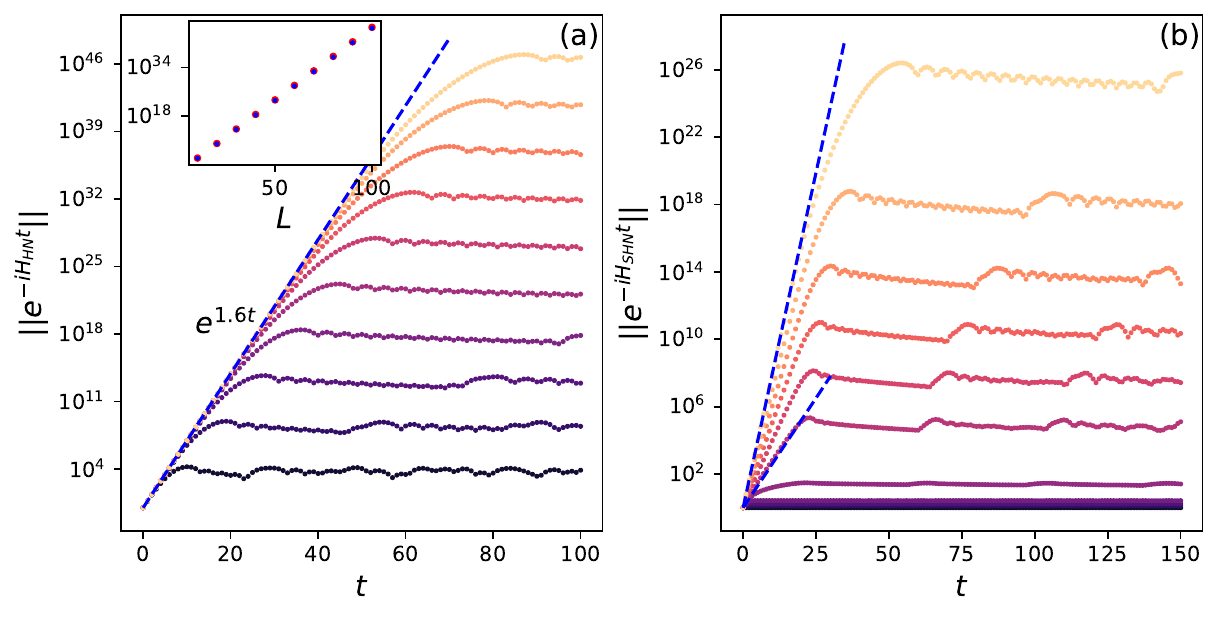}}
\caption{The evolution of $\lVert e^{-\mathrm{i}Ht}\rVert$ for Hatano-Nelson and symplectic Hatano-Nelson model. The numerical precision is set to be digit 50. (a) The Hatano-Nelson model, $\gamma=0.8$. From the darkest color to the lightest color, $L=10\sim100$. The blue dashed line is $e^{2\gamma t}$. In the inset, the red circles represent the condition number of the Hatano-Nelson model ($r^{L-1}$), while the blue dots represent the $\text{max}\lVert e^{-\mathrm{i}H_{\text{HN}}t} \rVert$. (b) The Symplectic Hatano-Nelson model, $\delta=0.4, L=40$, from the darkest color to the lightest color, $\gamma=0.0\sim1.0$. The blue dashed lines are $e^{2\sqrt{\gamma^2-\delta^2}t}$. }
\label{EvoMatrixNorm}
\end{figure}

Here, we provide an analysis of the numerical errors in the non-Hermitian evolution.  As demonstrated earlier, strong normality can induce spectral instability, and we now show that it can also lead to unconventional transient dynamics. For example, consider the evolution of the $\lVert e^{-\mathrm{i}Ht}\rVert$ under OBC. As shown in Fig. \ref{EvoMatrixNorm},  for the Hatano-Nelson model and symplectic Hatano-Nelson model in the skin regime ($\gamma>\delta$), $\lVert e^{-\mathrm{i}Ht}\rVert$ grows exponentially with time $t$ in the early stages. This behavior is peculiar because the energy spectra $\Lambda$ of both models are entirely real, and theoretically $\lVert e^{-\mathrm{i}\Lambda t}\rVert=1$. The discrepancy between $\lVert e^{-\mathrm{i}Ht}\rVert $ and $\lVert e^{-\mathrm{i}\Lambda t}\rVert$ originates from the strong non-normality. It is readily inferred that $\lVert e^{-\mathrm{i}Ht}\rVert\leq \lVert V\rVert\lVert V^{-1}\rVert\lVert e^{-\mathrm{i}\Lambda t}\rVert\leq \text{cond}(V)$, where the condition number provides an upper bound. Indeed, as shown in the inset of Fig. \ref{EvoMatrixNorm}(a), the red dots represent the theoretical values of the condition number ($r^{L-1}$), while the blue dots represent the maximum value of $\lVert e^{-\mathrm{i}H_{\text{HN}}t}\rVert$ during the evolution. Clearly, the condition number serves as a tight upper bound of $\lVert e^{-\mathrm{i}Ht}\rVert$. For the symplectic Hatano-Nelson model, the behavior of $\lVert e^{-\mathrm{i}H_{\text{SHN}}t}\rVert$ differs qualitatively for $\gamma>\delta$ and $\gamma<\delta$ as illustrated in Fig. \ref{EvoMatrixNorm}(b). This difference is attributed to the distinct behaviors of the condition number in these two regimes. 
Moreover, the initial growth of $\lVert e^{-\mathrm{i}Ht}\rVert$ follows an exponential form,  $e^{\alpha t}$, where $\alpha$ is determined by the numerical abscissa \cite{pseudospectra}. The numerical range of a matrix $H$ is defined to be $W(H)=\{\langle\psi|H|\psi\rangle|\ \lVert|\psi\rangle\rVert=1, |\psi\rangle\in\Bbb{C}^N \}$. For the Hatano-Nelson model, the numerical range is given by the PBC spectrum $E^{(m)}_{\text{PBC}}=2J\text{cos}k_m-\mathrm{i}2\gamma\text{sin}k_m$, which theoretically demands $\dfrac{d}{dt}\lVert e^{-\mathrm{i}H_{\text{HN}}t}\rVert_{t=0}\leq2\gamma$ \cite{pseudospectra}. This result is consistent with the fitting value $\alpha=2\gamma$ in Fig. \ref{EvoMatrixNorm}(a). Similarly, for the symplectic Hatano-Nelson model, the $\lVert e^{-\mathrm{i}H_{\text{SHN}}t}\rVert$ initially grows as $e^{2\sqrt{\gamma^2-\delta^2}t}$ as proved in Fig. \ref{EvoMatrixNorm}(b).

In addition, we also calculate the $\lVert e^{-\mathrm{i}ht}U\rVert$, where $h$ and $U$ are the matrix representations of the Hamiltonian and initial state, respectively (see Appendix. \ref{AppendixGaussianStateSimulation}). We observe that its behavior closely resembles that of $\lVert e^{-\mathrm{i}Ht}\rVert$. This indicates that the non-Hermitian evolution $\dfrac{e^{-\mathrm{i}Ht}|\psi_0\rangle} {\lVert e^{-\mathrm{i}Ht}|\psi_0\rangle\rVert}$ can be numerically challenging. Specifically speaking, as time $t$ increases, $\lVert e^{-\mathrm{i}Ht}|\psi_0\rangle\rVert$ exponentially grows until it approximately approaches the value of  $\text{cond}(V)$. For highly non-normal systems, $\text{cond}(V)$ can be extremely large, causing the elements of $\dfrac{e^{-\mathrm{i}Ht}|\psi_0\rangle}{\lVert e^{-\mathrm{i}Ht}|\psi_0\rangle\rVert}$ to become much smaller than the default double precision, leading to numerical errors. Furthermore, during the timescale in which $\lVert e^{-\mathrm{i}Ht}\psi_0\rVert$ is smaller than about $O(10^{16})$, we can predict that the non-Hermitian evolution is numerically correct with double precision. For example, as shown in Fig. \ref{CurrentEvoHN} and \ref{EEntropySymplecticHN}, the current and the entanglement entropy with double precision agree with the results of digit $50$ for an initial period. In summary, the large condition number not only induces the spectrum instability as illustrated in Sec. \ref{SecNumericalErrorDiag} but also contributes to the numerical instability of the non-Hermitian evolution.

\section{Dynamics requires higher simulation precision than spectrum}\label{SecMisunderstanding}
\begin{figure}[htb]
\centering
\subfigure{
\includegraphics[height=5.0cm,width=8.0cm]{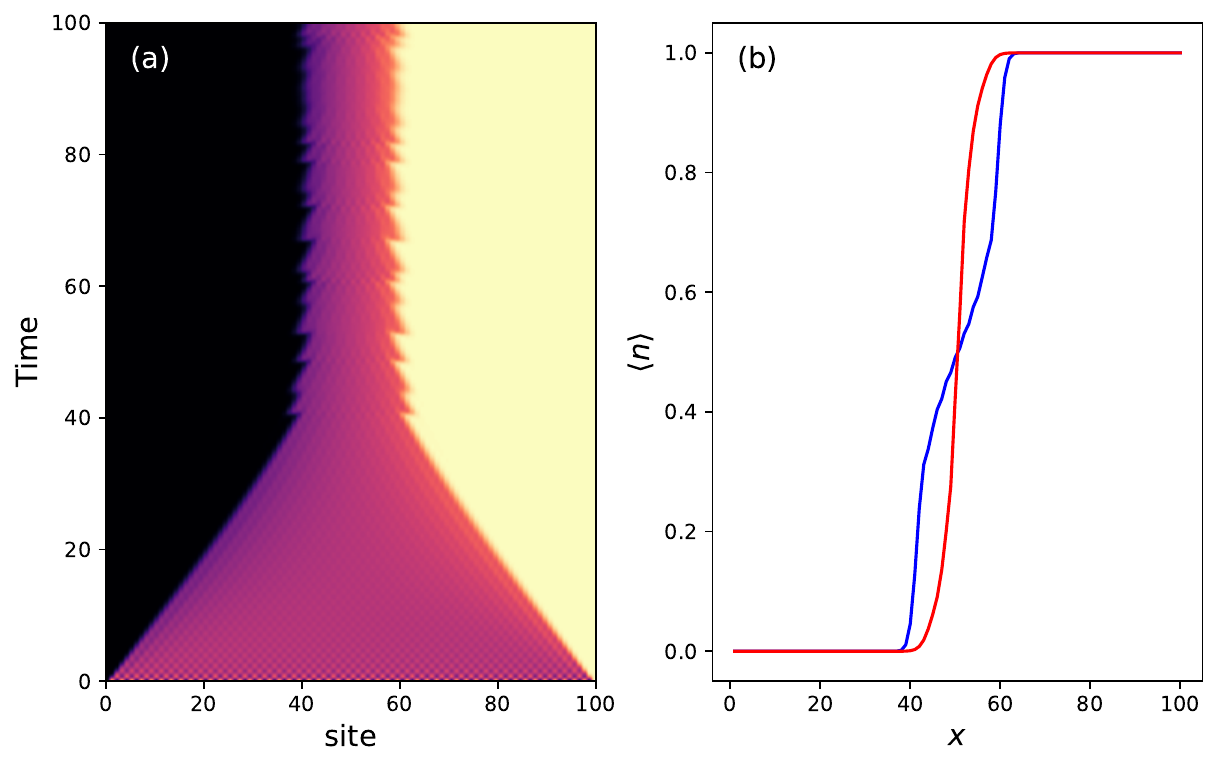}}
\caption{$L=100$, $\gamma=0.8$. (a) The density evolution of the Hatano-Nelson model under OBC with a physically incorrect half-filling middle region. The digit is $P=30$. (b) The steady-state density of digit 30 and digit 50.}
\label{P30L100}
\end{figure}

In practice, diagonalizing a non-normal matrix ($H$) is computationally much faster than calculating non-Hermitian evolution ($e^{-\mathrm{i}Ht}|\psi_0\rangle/\lVert e^{-\mathrm{i}Ht}|\psi_0\rangle\rVert$). This raises an important question: if we obtain a reliable spectrum with a precision of $P$ digits (i.e., numerical errors are smaller than a threshold $\epsilon$), does this guarantee that the non-Hermitian evolution computed with the same precision $P$ will also be reliable, with numerical errors of the same order? Our results suggest that the answer is no.
For example, for the Hatano-Nelson model with $L=100$, $J=1.0$, $\gamma=0.8$: as shown in Fig. \ref{NumericalDiag}(c), with  digits $P=30$, the diagonalization results are completely real with numerical errors $|E_j(P=30)-E^{(j)}_{\text{OBC}})| < 10^{-6}$, $\forall j$. However, the numerical errors of non-Hermitian evolution are far larger than the $10^{-6}$ as depicted in Fig. \ref{P30L100}, leading to qualitatively incorrect simulation results. As demonstrated in Fig. \ref{ConnectionSupp2}(a2), although the energy spectrum is reliable,  the eigenvectors still deviate from the analytical results. Specifically, as we claimed before, the theoretical values of wave functions near the left edge are too small, beyond the scope of the digit $30$.  
These numerical errors in eigenvectors can serve as a signature of the incorrectness of the non-Hermitian evolution. In other words, the unreliability of eigenvectors means that non-Hermitian evolution is also unreliable. 
This correspondence is particularly clear in the Hatano-Nelson model, where the absolute values of the wave functions behave approximately as $r^{x}$, and the condition number scales as $r^{L-1}$. Given that the absolute value of the wave function around the right edge is $O(1)$, the absolute value of the wave function near the left boundary is about $O(1)/r^L$. For the non-Hermitian evolution, it is known the condition number gives a tight upper bound of the $\lVert e^{-\mathrm{i}Ht}\rVert$, implying that the minimum value of $1/\lVert e^{-\mathrm{i}Ht}\rVert$ scales as $r^{-(L-1)}$. Therefore, the incorrectness of eigenvectors indicates that the non-Hermitian evolution will also be unreliable. In summary, obtaining a reliable spectrum alone is insufficient to ensure accurate non-Hermitian evolution. The accuracy of the eigenvectors plays a more crucial role in the numerical stability of non-Hermitian evolution. In other words, numerically correct evolution results require higher precision than numerically correct spectrum results.

Finally, while the models presented here are solvable and thus suitable for benchmarking, most non-Hermitian models, such as disordered, high-dimensional, and interacting systems, cannot be solved exactly. Consequently, it is generally challenging to assess the reliability of the spectrum and non-Hermitian evolution. However, our work underscores the importance of non-normality in determining the numerical stability of both the spectrum and non-Hermitian evolution. The strong non-normality is a universal property for non-Hermitian systems with NHSE, and it can be effectively described by the condition number.
Therefore, given a non-Hermitian Hamiltonian, we can begin by diagonalizing it for small system sizes to extract the relationship between the condition number and the system size. Then, we can extrapolate the system size to larger values and monitor the numerical magnitude of the condition number. 
If the condition number is on the order of $O(10^{N})$ and exceeds the double-precision limit of approximately $O(10^{16})$, higher numerical precision is required. To ensure the reliability of the results, one may select a working precision of $P = N + 20$ digits.
This consideration is particularly important for non-Hermitian systems exhibiting the NHSE, where the condition number typically grows exponentially with system size. It is therefore essential to estimate its magnitude before numerical computations. Appendix~\ref{AppendixDisorderInteraction} presents examples of disordered and interacting Hatano-Nelson models, both of which suffer from numerical instability. These cases highlight the generality of such instability in NHSE systems and underscore the importance of monitoring the condition number to ensure reliable numerical results.

\section{Conclusions and Discussions}\label{SecConclusion}
In this work, we investigate the numerical instability of non-Hermitian Hamiltonians with NHSE, a phenomenon often overlooked in practice \cite{NHSEEPTKawabata}. 
Our results demonstrate that numerical errors can significantly impact the spectra, wave functions, and non-Hermitian evolution. Mathematically, this numerical instability is attributed to the large condition number of the non-Hermitian Hamiltonian with NHSE, which grows exponentially with the system size. Our work demonstrates that a reliable spectrum alone is not sufficient to ensure accurate non-Hermitian evolution. Instead, it is crucial to verify the reliability of the wave functions. Therefore, it is paramount to estimate the condition number before numerical computations. Assuming the condition number is on the order of $O(10^N)$, we can set the precision digit to be $P=N+20$, serving as a practical guideline for choosing precision in simulations involving highly non-normal matrices.

However, increasing numerical precision significantly slows down calculations, as the process shifts from numerical to symbolic computation. Consequently, high-precision numerical calculations can be computationally expensive.
We have observed that the results $\gamma=-0.8$ are much more accurate than the $\gamma=0.8$ at double precision, despite both corresponding to the same non-Hermitian Hamiltonian after a transpose operation. This suggests that, in the future, it may be possible to develop more efficient algorithms that yield accurate results without relying on high-precision symbolic computations.

Moreover, the skin effect has also been studied in the Markovian open quantum systems. According to quantum trajectory theory \cite{quantumtrajectory}, the non-Hermitian evolution studied in this work can be understood as a special trajectory devoid of quantum jumps. With well-designed Lindblad operators, recent studies have explored the skin effect within the Lindbladian framework, incorporating all possible quantum trajectories. Remarkably, in such Lindbladian systems, the steady state remains approximately half-filled---i.e., localized on one side of the system---and the entanglement entropy obeys an area law for $\gamma\neq0$ \cite{MISE,MISELongrange,MISEDPT,MISEDelocalization,MISELiu,MISETilted}.
The dynamics of Lindbladians exhibiting the skin effect have also been simulated using Gaussian-state methods. Moreover, the condition number of the Lindbladian grows exponentially with system size. This observation raises the intriguing question of whether similar numerical instabilities arise in such open quantum systems.

In this work, we present one-dimensional free, disordered, and interacting Hatano–Nelson models to illustrate the generality of numerical instability. These findings suggest that such instabilities are likely to be widespread in non-Hermitian and Lindbladian systems exhibiting the skin effect, regardless of the presence of disorder, dimensionality, or interactions
\cite{DisorderNHSE,zhang2025scalefreelocalizationversusanderson,zhangUniversalNonHermitianSkin2022,Amoeba,DimensionTrans,xiong2024nonhermitianskineffectarbitrary,songfragilenonblochspectrum,ManyBodyNHSEgaugecoupling,NHSEBoseGas,SunBosonicHN,NHBoseHubbard,hamanakaInteractioninducedLiouvillianSkin,maoliouvillian2024,BoundarySensitiveLindb}. Finally, this extreme sensitivity is not only relevant for numerical simulations but also for experiments, which are inevitably affected by defects, disorder, or noise. We predict that such imperfections in experimental setups may also lead to results that deviate significantly from the clean limit.

\begin{acknowledgments}
The pseudospectra are calculated with the EigTool in MATLAB \cite{EigTool}.
X.F. and S.C. were supported by the National Key Research
and Development Program of China (Grant No. 2023YFA1406704), the NSFC under Grants No. 12474287 
and No. T2121001. SXZ acknowledges the support from a start-up grant at IOP-CAS.
\end{acknowledgments}

\clearpage
\appendix
\renewcommand{\theequation}{S\arabic{equation}}
\setcounter{equation}{0}
\renewcommand{\thefigure}{S\arabic{figure}}
\setcounter{figure}{0}

\begin{figure*}[htb]
\centering
\subfigure{
\includegraphics[height=6.0cm,width=18.0cm]{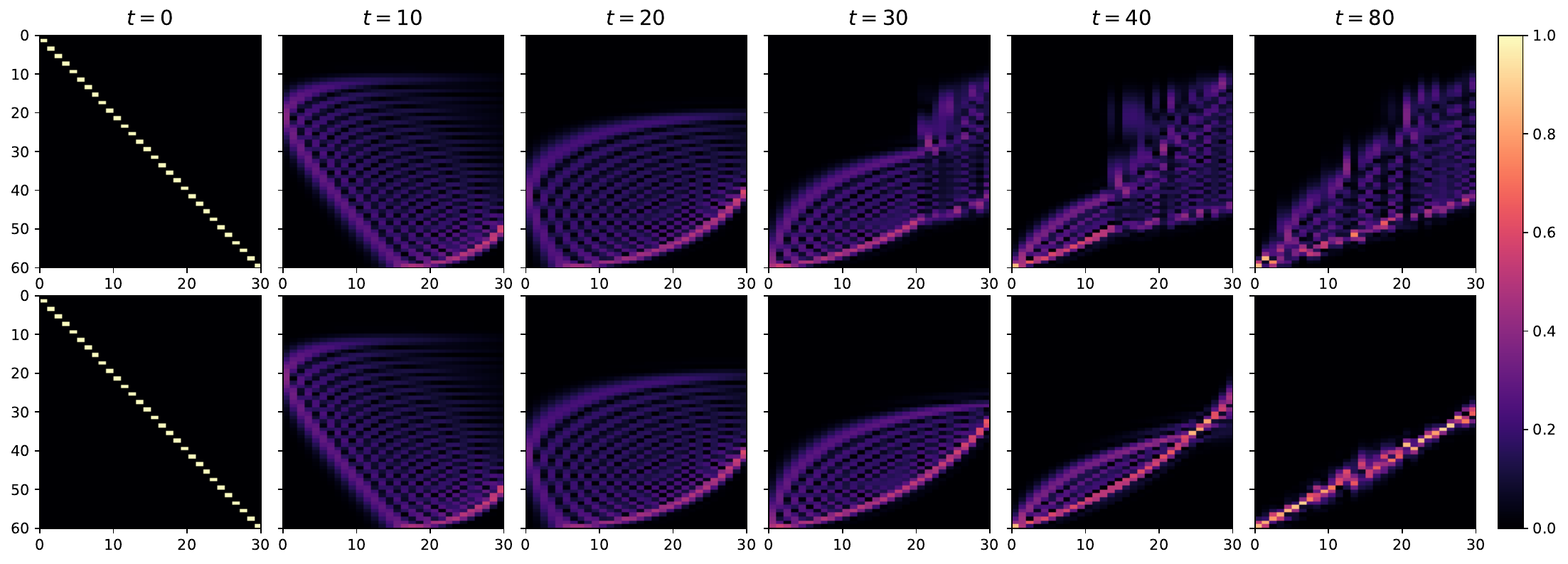}}
\caption{The evolution of the $|U|$ matrix ($|U|$ means elementwise absolute value) for the Hatano-Nelson model under OBC. The parameters are $L=60, \ \delta t=10$, $\gamma=0.8$.  The upper (lower) panels are double-precision (digit 50).}
\label{UEvoHN}
\end{figure*}

\section{Numerical asymmetry induced by $QR$ algorithm}\label{AppendixNumericalAsymmetry}
In the main text, we observed that double-precision numerical results for $\gamma=-0.8$ are significantly more accurate than those for $\gamma=0.8$, even though the corresponding Hamiltonians are the same after a transposition. To investigate the origin of this numerical asymmetry, we simulate the diagonalization procedure.
The LAPACK routines for unsymmetric matrices follow three main steps:
(1) Balancing: This step applies a similarity transformation: $\tilde{H}=D^{-1}P^{-1}HPD$, where $D$ is a diagonal scaling matrix and $P$ is a permutation matrix (if necessary).
(2) Hessenberg Reduction: The balanced matrix $\tilde{H}$ is reduced to upper Hessenberg form $H^{\prime}$, $H^{\prime}=Q^T\tilde{H}Q$, in which $Q$ is an orthogonal matrix. The goal is to concentrate most of the mass on the diagonal and first sub-diagonal.
(3) $QR$ algorithm: Eigenvalues are extracted from the Heisenberg matrix using the implicit double-shift $QR$ iteration. The goal is to find the real Schur form---a quasi-upper-triangular matrix from which real and complex eigenvalues are obtained from $1\times1$ and $2\times2$ blocks, respectively.

Numerical experiments show that step 3, the $QR$ algorithm, is the source of the numerical asymmetry.
In MATLAB, steps 1 and 2 correspond to the functions ``balance" and ``hess", respectively, while step 3. The step (3) can be executed using the function ``schur".
We consider the Hatano-Nelson model with parameters $L=40$ and $\gamma=\pm0.8$. Using MATLAB’s ``schur" function to perform the Schur decomposition, we observe that for $\gamma=0.8$, the resulting quasi-upper-triangular matrix contains many $2\times2$ blocks along the diagonal, indicating complex eigenvalues. In contrast, for $\gamma=-0.8$, the quasi-upper triangular matrix is composed of $1\times1$ blocks, corresponding to real eigenvalues. This observation confirms that the numerical asymmetry originates from the Schur decomposition.
To further monitor the occurrence of the numerical asymmetry, we simulate the $QR$ and $RQ$ iterations---key components of the Schur decomposition. After $4000$ iterations, both Hamiltonians are transformed into quasi-upper-triangular matrices. Consistently, we observe that $\gamma=0.8$ yields a spectrum dominated by $2\times2$ blocks, whereas $\gamma=-0.8$ yields $1\times1$ blocks, reproducing the asymmetry seen in the ``schur" output.

A natural question arises: Why does the $QR$ algorithm exhibit greater numerical stability for $\gamma=-0.8$ than for $\gamma=0.8$?

To address this, we examine the standard implementation of the $QR$ algorithm via Householder reflections. This method iteratively eliminates sub-diagonal elements column by column using a sequence of orthogonal transformations. At each step $k$, a Householder matrix $T_k$ is constructed to zero out the sub-diagonal elements of column $k$ of a $L\times L$ matrix $A$. The Householder reflector is given by
\begin{equation}
    T_k=I-2\dfrac{v_kv_k^T}{v_k^Tv_k},
\end{equation}
where $v_k$ is chosen so that $T_kA$ has zeros below the diagonal in column $k$. The overall orthogonal matrix is the product $Q=T_1T_2T_3\cdots T_{L-1}$, and the transformed matrix becomes upper triangular: 
$R=T_{L-1}T_{L-2}\cdots T_1A$.

We implement this $QR$ decomposition using Householder reflections with parameters $L=40$ and $\gamma=\pm0.8$. For clarity, we introduce the following notation. For $\gamma=0.8$, let: $H^{\text{pos}}=Q^{\text{pos}}_0R^{\text{pos}}_0$, $H_1^{\text{pos}}=R^{\text{pos}}_1Q^{\text{pos}}_1$, $H_1^{\text{pos}}=Q^{\text{pos}}_1R^{\text{pos}}_1$,  $H_2^{\text{pos}}=R^{\text{pos}}_1Q^{\text{pos}}_1$, and so on. Similarly, for $\gamma=-0.8$, let: $H^{\text{neg}}=Q^{\text{neg}}_0R^{\text{neg}}_0$, $H_1^{\text{neg}}=R^{\text{neg}}_1Q^{\text{neg}}_1$, $H_1^{\text{neg}}=Q^{\text{neg}}_1R^{\text{neg}}_1$, $H_2^{\text{neg}}=R^{\text{neg}}_1Q^{\text{neg}}_1$, and so on.
To assess numerical stability, we perform $QR$ and $RQ$ iterations using both double-precision and digit $50$. For $\gamma=-0.8$, the matrices $H_j^{\text{neg}}$ agree closely between the two precisions, indicating numerical stability. In contrast, for $\gamma=0.8$, significant deviations arise in $H_j^{\text{pos}}$ after only a few iterations, indicating numerical insufficiency of double precision. 
This discrepancy suggests that $H_j^{\text{neg}}$ is inherently more stable under Householder reflections than $H_j^{\text{pos}}$. Indeed, we observe a structural difference between them. The non-zero elements in the last few columns of $H_j^{\text{neg}}$ tend to be concentrated in the lower-right triangular region, whereas for $H_j^{\text{pos}}$, they concentrate in the upper-right region. This distinction affects the stability of the Householder transformation. In particular, while computing $T_k$ for $H_j^{\text{pos}}$, we can encounter extremely small denominators $v_k^Tv_k$ (small than $O(10^{-16})$), leading to numerical instability. Such small denominators are not observed for $H_j^{\text{neg}}$, which explains its enhanced numerical robustness.

In summary, our numerical experiments demonstrate that the asymmetry in the numerical stability between $\gamma=\pm0.8$ arises from structural differences in the matrices generated during $QR$ and $RQ$ iterations. The matrices $H_j^{\text{pos}}$ associated with $\gamma=0.8$ are more prone to instability due to the emergence of extremely small denominators in the Householder reflection process. In contrast, the structure of $H_j^{\text{neg}}$ for $\gamma=-0.8$ avoids this issue, resulting in greater numerical stability.

\section{Gaussian state simulation method and observables}
\label{AppendixGaussianStateSimulation}
The Hatano-Nelson and the symplectic Hatano-Nelson model discussed in the main text are non-interacting and $U(1)$ symmetric, meaning they are quadratic. If the initial state is Gaussian, the evolved state will preserve the Gaussianity through the non-Hermitian evolution. The Gaussian state simulation method has been illustrated in detail in Ref. \cite{NHSEEPTKawabata}; for completeness, we briefly outline it below.
The free-fermionic Gaussian state can be written as 
\begin{equation}
|\psi(t)\rangle=\prod_{n=1}^{N}\left(\sum_{j=1}^{L}U_{j,n}(t)c_{j}^{\dagger} \right)|0\rangle,
\end{equation}
in which $|0\rangle$ is the fermionic vacuum state.
$|\psi(t)\rangle$ is a Slater determinant state of
$N$ fermions, with the columns of $U$ encoding the single-particle wave functions. Therefore, the many-body state $|\psi\rangle$ can be represented by the $L\times N$ matrix $U$, which satisfy $U^{\dagger}U=I$. The time evolution of the $|\psi(t)\rangle$ can also be reduced to the evolution of the matrix $U$: 
\begin{equation}
\begin{aligned}
|\psi(t+\delta t)\rangle&= e^{-\mathrm{i}H\delta t}|\psi(t)\rangle \\
&=\prod_{n=1}^N\left(\sum_{j=1}^L U_{j,n}(t) e^{-\mathrm{i}H\delta t}c_j^{\dagger}e^{\mathrm{i}H\delta t} \right)|0\rangle \\ 
&=\prod_{n=1}^N\left(\sum_{j=1}^L U_{j,n}(t)\sum_{m=1}^L[e^{-\mathrm{i}h\delta t}]_{m,j}c_m^{\dagger} \right)|0\rangle \\
&=\prod_{n=1}^N\left(\sum_{m=1}^L[e^{-\mathrm{i}h\delta t}U]_{m,n}c_m^{\dagger} \right)|0\rangle,
\end{aligned}
\end{equation}
in which we assume the $H=\sum_{i,j}h_{i,j}c^{\dagger}_ic_j$ and use the Baker-Campbell-Hausdorff formula $e^{-\mathrm{i}H\delta t}c^{\dagger}_j e^{\mathrm{i}H\delta t}=\sum_{m=1}^L[e^{-\mathrm{i}h\delta t}]_{m,j}c^{\dagger}_m$.
Therefore, the non-Hermitian evolution simplifies to $U(t+\delta t)=e^{-\mathrm{i}h\delta t}U(t)$. Noticing the denominator $\lVert e^{-\mathrm{i}Ht}|\psi(t)\rangle\lVert$, we should perform normalization after each time step. The normalization can be achieved by performing $QR$ decomposition $e^{-\mathrm{i}h\delta t}U(t)=QR$, then reassign $U(t+\delta t)=Q$. 

The matrix $U(t)$ encapsulates all quantum dynamical information. 
Once its evolution is determined, various observables—including correlation functions, density distributions, currents, and entanglement entropy—can be computed. Firstly, the correlation matrix is given by 
\begin{equation}
C_{i,j}(t):=\langle\psi(t)|c^{\dagger}_ic_j|\psi(t)\rangle=[U(t)U^{\dagger}(t)]_{j,i}.
\end{equation}
The local particle number is extracted from the diagonal element of the correlation matrix $ n_{j} (t)=C_{j,j}(t)$. Moreover, we consider the local current and total current. The local current between site $j$ and $j+1$ is defined as $I_{j}=\langle\psi(t)|\dfrac{\mathrm{i}J}{2}(c^{\dagger}_{j}c_{j+1}-c^{\dagger}_{j+1}c_j)|\psi(t)\rangle$, in which $J$ is set to be $1$. The total current is $I(t)=\sum_{j=1}^{L-1}I_{j}(t)$. The half-chain von Neumann entanglement entropy $S(L,L/2)$ is given by 
\begin{equation}
   S(L,L/2)=-\sum_{j=1}^{L/2}\left[\lambda_j\ \ln\lambda_j+(1-\lambda_j)\ \ln(1-\lambda_j)\right], 
\end{equation}
where $\lambda_j$'s ($j=1,2,...L/2$) are the eigenvalues of the submatrix $[C]_{i,j=1}^{L/2}$.

For the symplectic Hatano-Nelson model, which consists of two coupled chains, $A$ and $B$, we adjust the $U$ matrix as  
\begin{equation}
|\psi(t)\rangle=\prod_{n=1}^N\left(\sum_{j=1}^L\sum_{s=A,B} U_{(j,s),n}c^{\dagger}_{j,s}\right)|0\rangle,
\end{equation}
in which $U$ is a $2L\times N$ matrix. Similarly, we can also calculate the correlation, density, current, and entanglement entropy. Importantly, we evaluate the currents separately for chains $A$ and $B$: $I_{s}=\langle\psi(t)|\dfrac{\mathrm{i}J}{2}(c^{\dagger}_{j,s}c_{j+1,s}-c^{\dagger}_{j+1,s}c_{j,s})|\psi(t)\rangle$, $s=A, B$.

In this paper, unless stated otherwise, we set the time step $\delta t=0.05$ and initialize the system in a N\'eel state. Additionally, we benchmark the Gaussian state method against exact diagonalization for small system sizes, finding excellent agreement (results not shown).

\section{Direct observation of the numerical errors}\label{AppendixUEvo}
Here, we directly observe the occurrence of numerical errors in non-Hermitian evolution, aiming to understand the origin of the spurious middle region.
The Gaussian-state simulation method consists of two main steps: acting on the operator $e^{-\mathrm{i}H}$ (more strictly $e^{-\mathrm{i}H\delta t}$) and subsequent $QR$ decomposition. Notably, the incorrect smooth middle region persists regardless of the time step $\delta t$. Therefore, for simplicity, we choose the parameters $\gamma=0.8,\ L=60,\ \delta t=10$, and compare results obtained using double precision and $50$-digit precision. It is known that the $j$-th column of the $U$ matrix represents the single-particle wave function of the $j$-th particle. The Fig. \ref{UEvoHN} implies that the particles will tend to localize at the right side. More importantly,  the steady-state $U$ matrix with double precision markedly deviates from the results with digit $50$. Specifically speaking, the correct results of digit $50$ show that the $j$-th particle has a high probability to localize around the site $L-j+1$, leading to a half-side occupied configuration. On the contrary, the $U$ matrix of double precision fails to converge properly, allowing particles to localize on the left side, thus inducing the smooth middle region. As stated in the Appendix. \ref{AppendixGaussianStateSimulation}, all the observables, including correlation functions, density distributions, currents, and entanglement entropy, are computed from $U$. Consequently, we argue that the double-precision numerical results reported in Ref. \cite{NHSEEPTKawabata} are unreliable whenever the condition number for the chosen Hamiltonian parameters exceeds $10^{16}$.

\begin{figure*}[htb]
\centering
\subfigure{
\includegraphics[height=6.56cm,width=17.5cm]{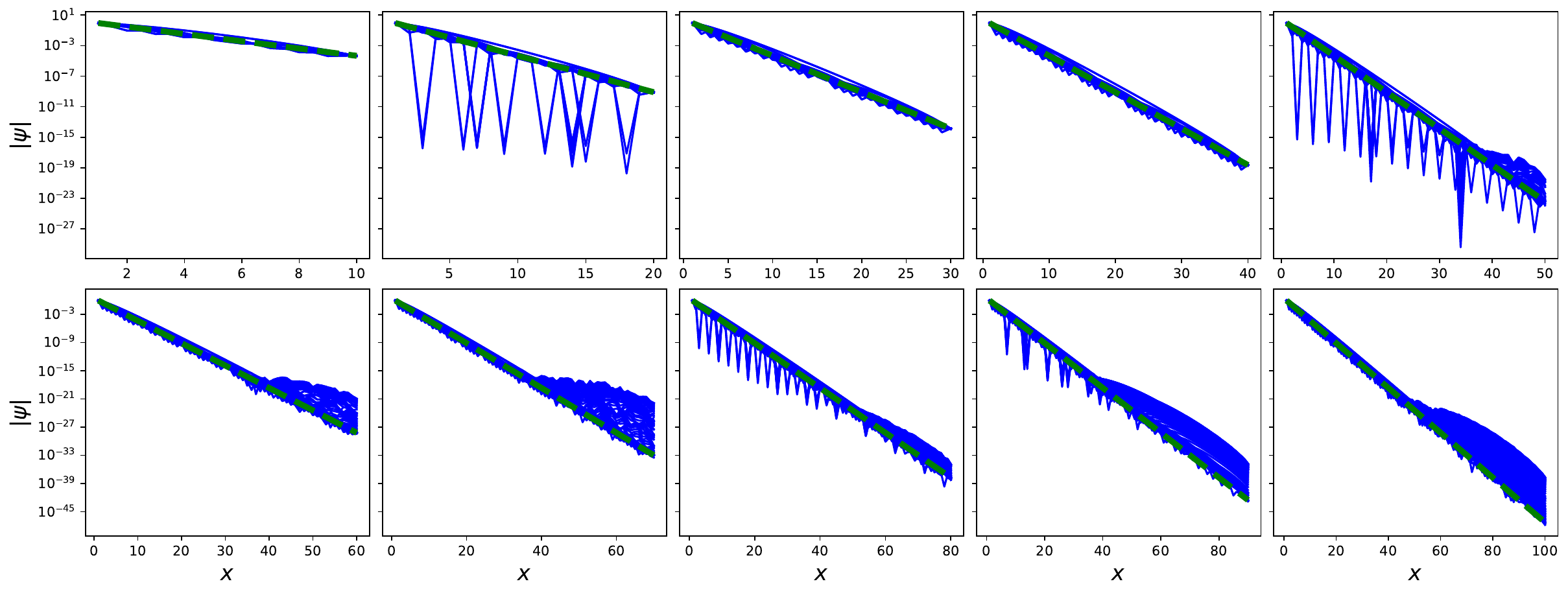}}
\caption{The numerical diagonalized wave functions of the Hatano-Nelson with different system sizes. All wave functions are plotted: the blue lines represent double-precision numerical results for $\gamma = -0.8$, while the dashed green lines indicate the theoretical exponential localization predicted by the model, given by $|\psi(x)| = ce^{-x/\xi}$, where the localization length is $\xi = 1/\ln\sqrt{(J+\gamma)/(J-\gamma)}$ and the prefactor $c$ depends on the system size. It should be noted that the dashed green lines are not the exact results of wave functions; they only approximately capture the exponential localization behaviors of the wave functions.
It is evident that the double-precision results of $\gamma=-0.8$ are much more accurate than $\gamma=0.8$ in Fig. \ref{NumericalDiagEigenVctors}. } 
\label{WavefunctionMinusGam}
\end{figure*}

\begin{figure}[htb]
\centering
\subfigure{
\includegraphics[height=4.0cm,width=9.0cm]{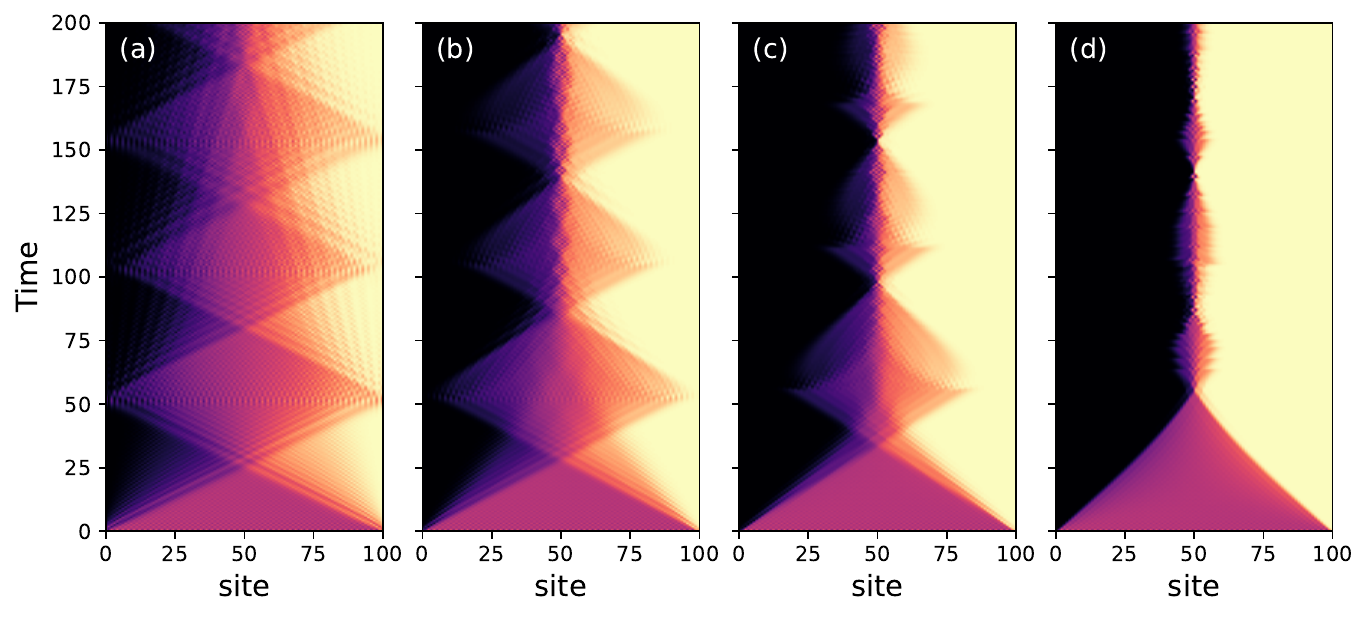}}
\caption{The density evolution of the Hatano-Nelson model under OBC. The precision is digit $50$, $L=100$. (a) $\gamma=0.1$, (b) $\gamma=0.3$, (c) $\gamma=0.5$, (d) $\gamma=0.8$.}
\label{DensityEvoGamRange}
\end{figure}

\begin{figure}[htb]
\centering
\subfigure{
\includegraphics[height=6.3cm,width=8.4cm]{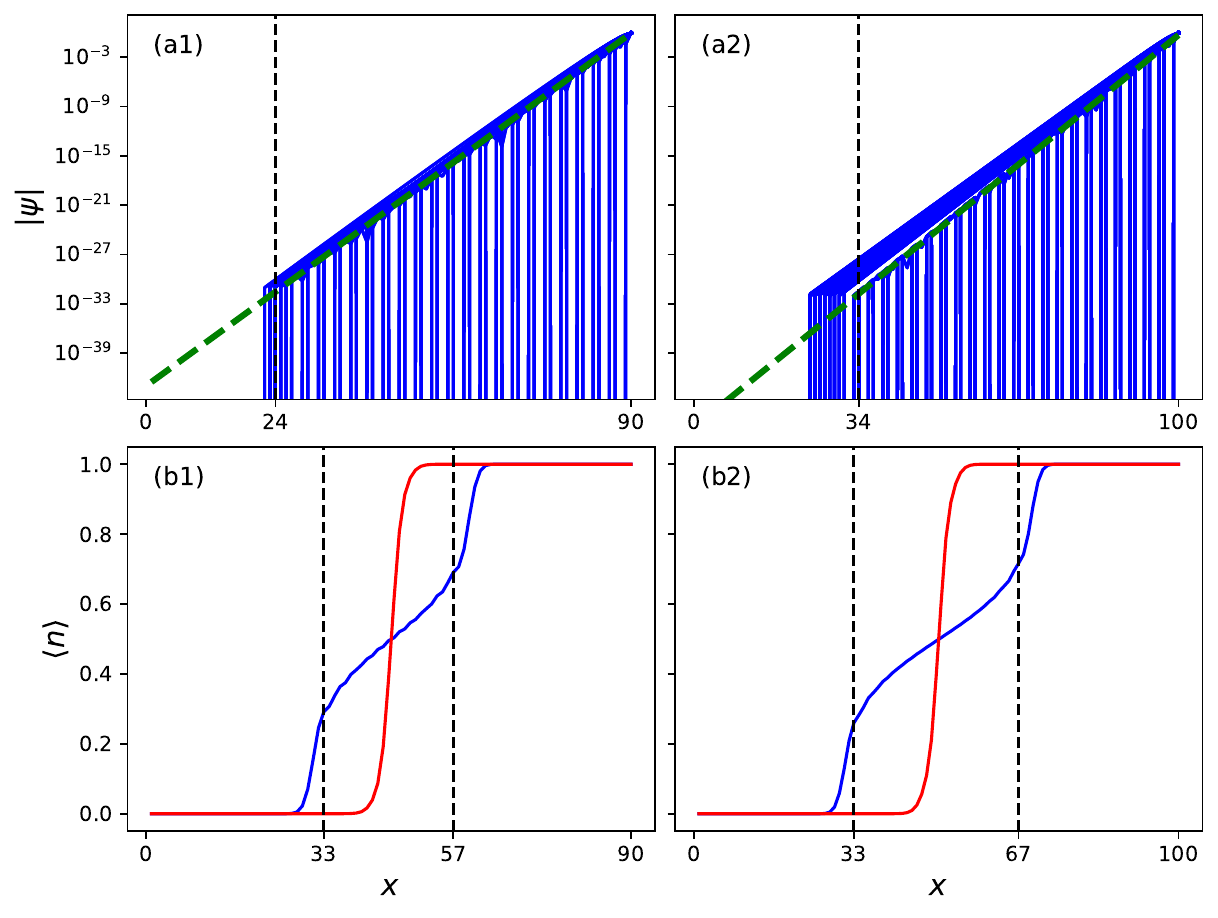}}
\caption{The connection between the width of the incorrect smooth region in the middle and the width of the incorrect region of the single-particle wave function with digit $20$. The Hatano-Nelson model under OBC with $\gamma=0.8$. (a1) and (a2) plot all the eigenvectors with digit $20$, $L=90, \ 100$. The green dashed line represents the theoretical exponential decay behavior. (b1) and (b2) plot the steady-state density distribution. The blue (red) line is the result of digit $20$ ($50$). }
\label{ConnectionSupp1}
\end{figure}

\begin{figure}[htb]
\centering
\subfigure{
\includegraphics[height=6.3cm,width=8.4cm]{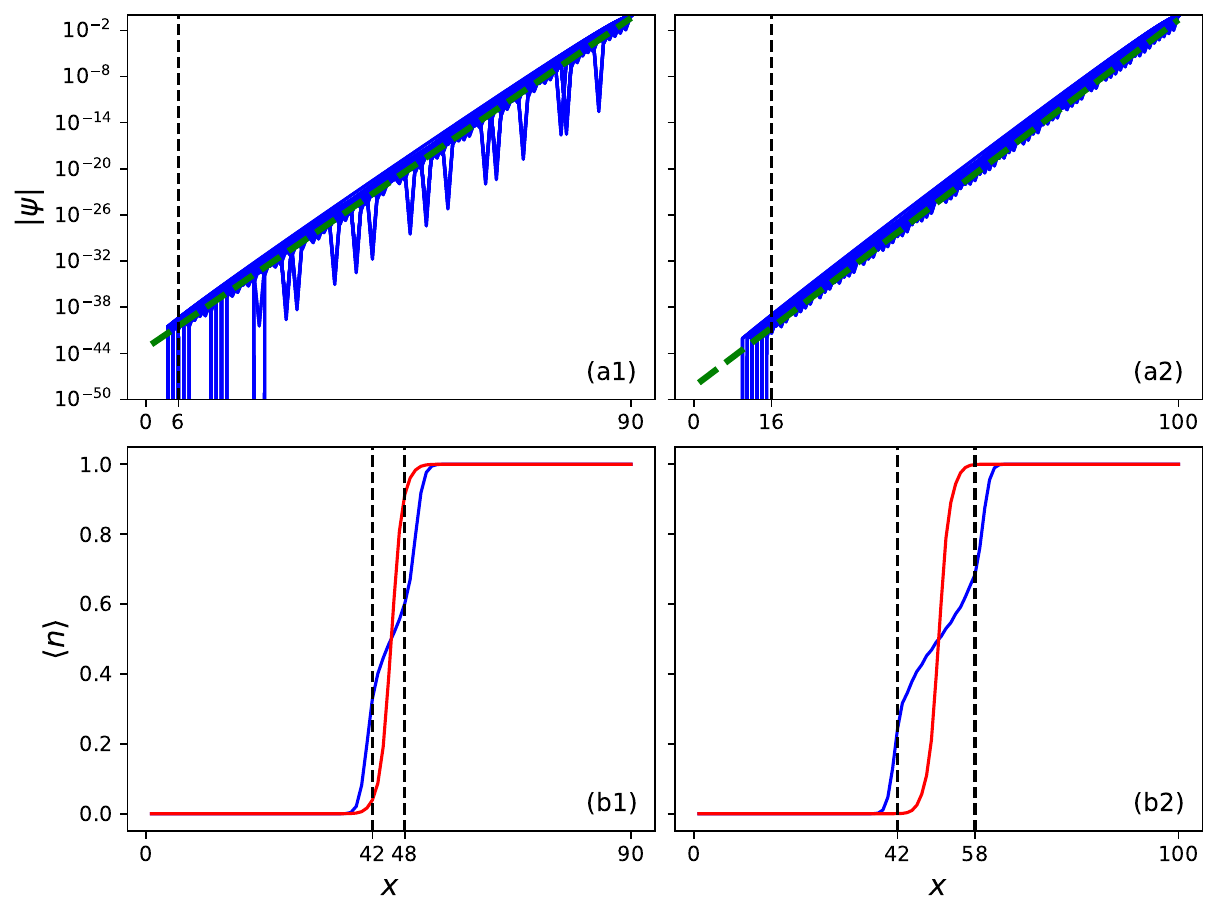}}
\caption{The connection between the width of the incorrect smooth region in the middle and the width of the incorrect region of the single-particle wave function with digit $30$. The Hatano-Nelson model under OBC with $\gamma=0.8$. (a1) and (a2) plot all the eigenvectors with digit $30$, $L=90, \ 100$. The green dashed line represents the theoretical exponential decay behavior. (b1) and (b2) plot the steady-state density distribution. The blue (red) line is the result of digit $30$ ($50$). }
\label{ConnectionSupp2}
\end{figure}

\section{Additional numerical results}\label{AppendixExtraNumericalRes}
As shown in Fig. \ref{DensityEvoGamRange}, 
for $\gamma\neq0$, the dynamics display the skin effect. We hypothesize that the steady state for any nonzero $\gamma$ should be close to one half-side occupied and another half-side unoccupied. For small $\gamma$, the oscillation amplitude is large and decays slowly, meaning the system takes a long time to reach the steady state. In contrast, for large $\gamma$, the system quickly approaches the steady state.

In the main text, we discuss the potential connection between the numerical errors in the wave functions and the errors in non-Hermitian evolution computed with double precision. Here, we present additional numerical data in Fig. \ref{ConnectionSupp1} and Fig. \ref{ConnectionSupp2} to further substantiate this observation, highlighting the correlation between the spatial extent of errors in the single-particle wave functions and in non-Hermitian evolution computed with the specified precision (digits $20$ and $30$).

\section{Numerical instability of non-Hermitian disordered and interacting systems}\label{AppendixDisorderInteraction}
\begin{figure}[htb]
\centering
\subfigure{
\includegraphics[height=4.2cm,width=8.4cm]{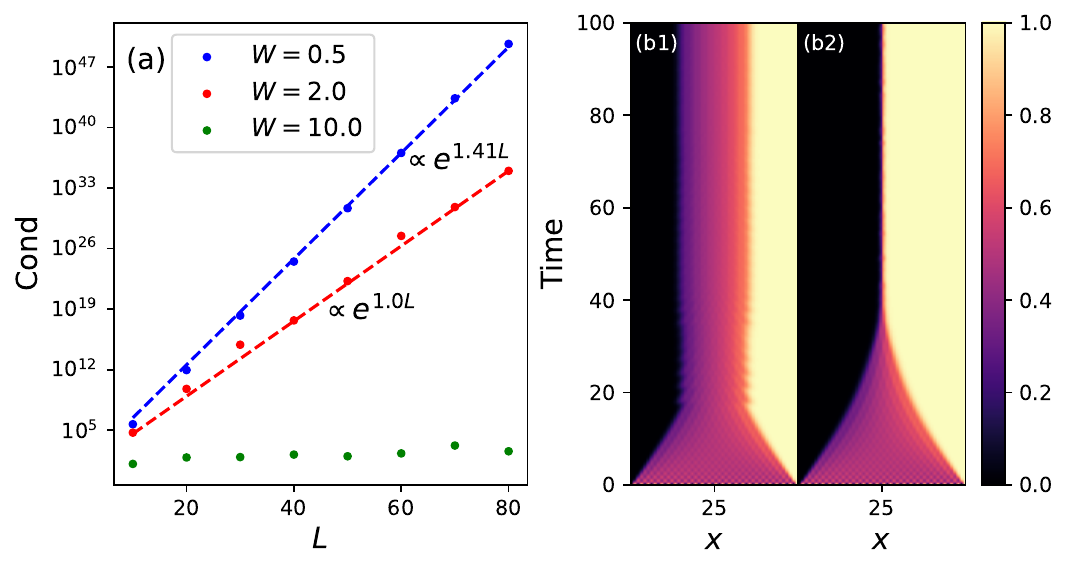}}
\caption{The disordered Hatano-Nelson model under OBC. $\gamma=0.9$, the data is averaged over 50 samples. (a) The condition number for different disorder strengths and system sizes. (b1) The density evolution with double precision.  $W=0.5$, $L=50$. (b2) The density evolution with digit 40. $W=0.5$, $L=50$.}
\label{Disorder}
\end{figure}

We further investigate the disordered and interacting Hatano-Nelson models to demonstrate the generality of the observed numerical instability.
The disordered Hatano-Nelson model is given by 
\begin{equation}
H_{\text{dis}}=\sum_{j}[(J+\gamma)c^{\dagger}_{j+1}c_j+(J-\gamma)c_j^{\dagger}c_{j+1}+\omega_jc_j^{\dagger}c_j],
\end{equation}
where $w_j$ is a random on-site potential drawn from a uniform distribution $\omega_j\in[-W, W]$. Under open boundary conditions (OBC), the spectrum of this disordered Hatano-Nelson model remains real, as the Hamiltonian can be transformed into a Hermitian form via a similarity transformation $Q$, as introduced in Eq. \eqref{Eq2}. 
It is well known that wavefunctions of the disordered Hatano-Nelson model transition from skin localization to Anderson localization with the increase in disorder strength $W$ \cite{DisorderNHSE,pseudospectraNHSE2}. As shown in Fig. \ref{Disorder}, the numerical instability persists in the weak disorder regime, where the condition number grows exponentially with the system size. On the contrary, the strong disorder leads to Anderson localization, and the numerical instability is eliminated due to the small condition number \cite{pseudospectraNHSE2}. For example,
in Fig. \ref{Disorder}(b1) and (b2), choosing parameters $W=0.5$, 
$L=50$, the condition number is much larger than $O(10^{16})$, so the density evolution displays the numerical instability as expected. Furthermore, we have also checked that the corresponding computed spectrum and wavefunctions are numerically unstable (not shown).

\begin{figure}[htb]
\centering
\subfigure{
\includegraphics[height=3.0cm,width=8.4cm]{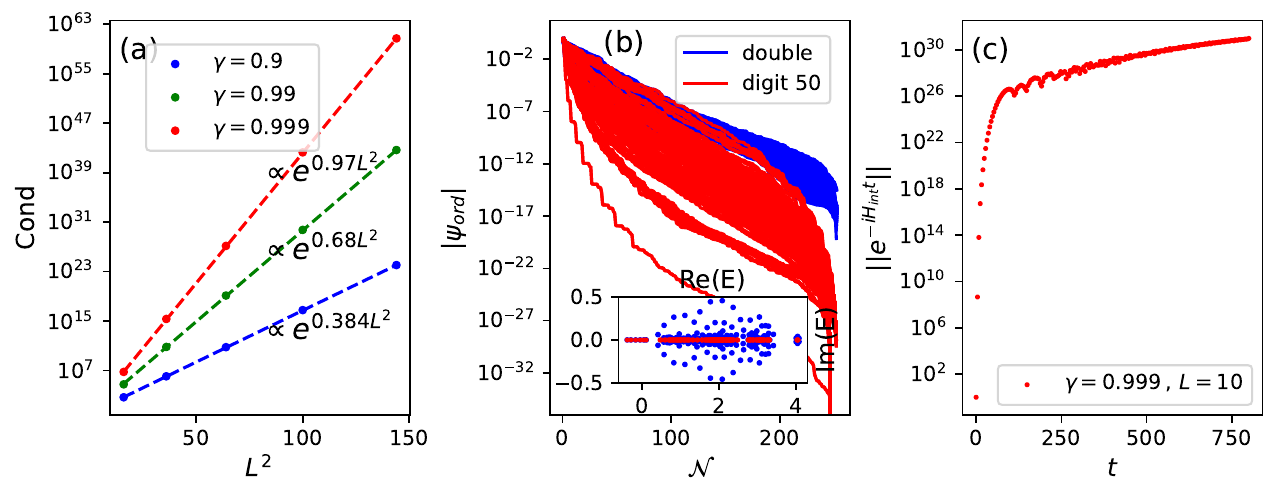}}
\caption{The half-filled interacting Hatano-Nelson model under OBC. (a) The condition number for different non-reciprocity $\gamma$ and system sizes $L$. (b) The main plot depicts the absolute values of all the wavefunctions, which are rearranged from the largest to the smallest for better comparison. The $x$ axis of the main plot is the index of the Fock basis. The parameters are $\gamma=0.99$, $L=10$, particle number $N=5$. The size of the Hilbert space is $C_{L}^N=252$. \sxz{no mention on the inset} (c) The evolution of the norm $\lVert e^{-\mathrm{i}H_{\text{int}}t}\rVert$. $\gamma=0.999$, $L=10$, digit 50.}
\label{Interacting}
\end{figure}

The interacting Hatano-Nelson model is defined as  
\begin{equation}
H_{\text{int}}=\sum_j[(J+\gamma)c_{j+1}^{\dagger}c_j+(J-\gamma)c^{\dagger}_jc_{j+1}+U_{\text{int}}n_jn_{j+1}],
\end{equation}
in which $U_{\text{int}}$ denotes the nearest-neighbor interaction.
The spectrum of the interacting Hatano-Nelson model remains real, since the model can also be mapped into a Hermitian Hamiltonian through a similarity transformation $Q_{\text{int}}=\text{exp}(\sum_{j}gjn_j)$ with $e^g=r=\sqrt{(J+\gamma)/(J-\gamma)}$. This transformation yields
\begin{equation}
Q_{\text{int}}^{-1}H_{\text{int}}Q_{\text{int}}=\sum_j\sqrt{J^2-\gamma^2}(c_j^{\dagger}c_{j+1}+\text{h.c.})+U_{\text{int}}n_jn_{j+1}. 
\end{equation}
As shown in Fig. \ref{Interacting}(a), the condition number for the half-filled case scales exponentially with $L^2$.
In particular, for the extreme non-reciprocity cases, i.e., $\gamma$ close to $1$, although the system size is limited for exact diagonalization, the condition number can still exceed $O(10^{16})$. Therefore, the default double precision becomes insufficient, resulting in inaccurate spectra and wavefunctions, as illustrated in Fig. \ref{Interacting}(b).
Moreover, as shown in Fig. \ref{Interacting}(c), the norm $\lVert e^{-\mathrm{i}H_{\text{int}}t}\rVert$ can be much larger than $O(10^{16})$, so the non-Hermitian evolution will also be numerically incorrect with double precision.

In conclusion, numerical instability is a general feature of systems exhibiting the non-Hermitian skin effect (NHSE), regardless of the presence of disorder or interactions. The key underlying issue is the exponential scaling of the condition number with system size. Therefore, it is essential to estimate the condition number for small systems and extrapolate it to larger sizes to determine the required numerical precision for reliable computation.

\bibliographystyle{apsreve}
\bibliography{ref}

\end{CJK*}
\end{document}